\documentclass[usenatbib]{mn2e}

\usepackage{mathtools}
\usepackage{graphicx}
\usepackage{epstopdf}
\usepackage{natbib}
\usepackage{verbatim}

\newcommand{\apj}{ApJ} \newcommand{\apjl}{ApJL}
\newcommand{\aap}{A\&A} \newcommand{\araa}{Ann. Rev. Astr. Ap.}
 \newcommand{\apjs}{ApJS} 
\newcommand{\mnras}{MNRAS} \newcommand{\nat}{Nature}
\newcommand{\pasp}{PASP} 

\newcommand{\rmxaa}{Rev. Mexicana Astron. Astrofis.}

 \usepackage{times}
\renewcommand{\thefootnote}{\fnsymbol{footnote}}

\def\lsim{\mathrel{\rlap{\lower4pt\hbox{\hskip1pt$\sim$}}
    \raise1pt\hbox{$<$}}}                
\def\gsim{\mathrel{\rlap{\lower4pt\hbox{\hskip1pt$\sim$}}
    \raise1pt\hbox{$>$}}}                

\begin{document}

\title[Extreme Ultraviolet Emission Lines at High Redshift]
{Ultraviolet Emission Lines in Young Low Mass Galaxies at $z\simeq 2$: Physical Properties and Implications for Studies at $z > 7$}

\author[Stark et al.] 
{Daniel P. Stark$^{1}$\footnote{1}, 
Johan Richard$^2$, 
Brian Siana$^3$, 
St\'{e}phane Charlot$^4$, 
William R. Freeman$^3$, \newauthor
Julia Gutkin $^4$,
Aida Wofford $^4$,
Brant Robertson$^1$,
Rahman Amanullah$^5$,
Darach Watson$^6$, \newauthor
\& Bo Milvang-Jensen$^6$ \\
 \\
$^{1}$ Steward Observatory, University of Arizona, 933 N Cherry Ave, Tucson, AZ, 85721, USA \\  
$^{2}$ Centre de Recherche Astrophysique de Lyon, Universite Lyon 1, 9 Avenue Charles Andre, 69561,  France  \\
$^{3}$ Department of Physics and Astronomy, University of California, Riverside CA 92521, USA \\
$^{4}$ UPMC-CNRS, UMR7095, Institut dÕAstrophysique de Paris, F-75014, Paris, France, 98 bis Boulevard Arago, F75014 Paris, France \\
$^{5}$ Department of Physics, Stockholm University, Albanova University Centre, SE 106-91 Stockholm, Sweden \\ 
$^{6}$ Dark Cosmology Centre, Niels Bohr Institute, University of Copenhagen, Juliane Maries Vej 30, 2100, Copenhagen \\
}
\date{Accepted ... ;  Received ... ; in original form ...}

\pagerange{\pageref{firstpage}--\pageref{lastpage}} \pubyear{2013}

\hsize=6truein
\maketitle

\label{firstpage}

\begin{abstract} 

We present deep spectroscopy of 17 very low mass (M$^\star \simeq 2.0\times10^6$ M$_\odot$  
to 1.4$\times$10$^9$ M$_\odot$) and low luminosity (M$_{\rm{UV}} \simeq -13.7$ to $-19.9$) 
gravitationally lensed galaxies in the redshift range $z\simeq 1.5-3.0$.   Deep rest-frame 
ultraviolet spectra reveal large equivalent width emission from numerous 
emission lines (NIV], OIII], CIV, Si III], CIII]) which are rarely seen in individual spectra of more massive star forming 
galaxies.     CIII]  is detected in 16 of 17 low 
mass star forming systems with rest-frame equivalent widths  as large as 13.5~\AA.    
Nebular CIV emission is present in the most extreme CIII] emitters, requiring an ionising source capable 
of producing a substantial component of photons with energies in excess of 47.9 eV.   
Photoionisation models support a picture whereby 
the large equivalent widths are driven by the increased electron temperature and enhanced ionising output 
arising from metal poor gas and stars (0.04 Z$_\odot$ to 0.13 Z$_\odot$), young stellar populations (6 Myr to 50 Myr), and 
large ionization parameters (log U = $-$2.16 to $-$1.84).   The young ages implied by  
the emission lines and continuum SEDs indicate that the extreme line emitters in our sample 
are in the midst of a significant upturn in their star formation activity.   The low stellar masses, blue UV 
colours, and large sSFRs of  our sample are similar to those of typical $z\gsim 6$ galaxies.   
Given the strong attenuation of Ly$\alpha$ in 
$z\gsim 6$ galaxies, we suggest that CIII] is likely to provide our best probe of early star forming galaxies 
with ground-based spectrographs and one off the most efficient means of confirming $z\gsim 10$ galaxies 
with the James Webb Space Telescope. 
\end{abstract} 


\renewcommand{\thefootnote}{\fnsymbol{footnote}}
\footnotetext[1]{E-mail: dpstark@email.arizona.edu}

\section{Introduction}
\label{sec:intro}

Low luminosity galaxies play an important role at high redshift.   Measurements of the UV luminosity function at 
$z\simeq 2$ indicate that more than 90\% of the total UV luminosity density comes from sub-L$^\star$ galaxies 
(e.g., Reddy \& Steidel 2009, Oesch et al. 2010, Alavi et al. 2014), and the dominance of such low luminosity systems becomes 
even more pronounced at yet earlier times (e.g., Bunker et al. 2010, Oesch et al. 2012, Bouwens et al. 2012, 
Schenker et al. 2013a, McLure et al. 2013).  The extent to which early galaxies pollute and ionise the intergalactic medium 
(IGM) depends sensitively on how efficiently baryons are converted to stars in the low mass dark matter halos 
thought to host the UV-faint population.  But while considerable progress has been achieved in our understanding 
of star formation and feedback in bright L$^\star$ galaxies at $z\simeq 2-3$ over the past decade (see Shapley 
2011 for a review), much less is known about the nature of low luminosity systems.

Galaxy formation may well proceed very differently in this low luminosity population.  The fraction of 
baryons locked in stars is thought to drop rapidly for dark matter halos less massive than 
$\sim 6\times$10$^{11}$ M$_\odot$ (e.g., Conroy \& Wechsler 2009, Moster et al. 2010, Guo et al. 2010, Behroozi et al. 2013), 
likely implying that the efficiency of star formation and gas cooling is greatly reduced in low mass halos.    
This deficiency is typically attributed to a combination of strong stellar feedback and photo-heating from the 
UV background (e.g., Larson 1974, Dekel \& Silk 1986,  Efstathiou 1992, Murray et al. 2005).    When feedback 
is strong in low mass halos, star formation histories are generally found to be `bursty' (e.g., Stinson et al. 2007, Shen et al. 2013, 
Teyssier et al. 2013, Hopkins et al. 2013).    The  fluctuations in star formation are predicted to occur  
on a dynamical timescale and may potentially play a fundamental role in modifying the dark matter 
distribution in dwarf galaxies 
(e.g., Pontzen \& Governato 2012, Teyssier et al. 2013).   

If star formation is indeed very bursty in low mass high redshift galaxies, then we would expect 
their spectra to look very different from more massive L$^\star$ galaxies that have been 
studied in great detail (e.g., Shapley et al. 2003) at the same cosmic epoch.   In the past several years, 
the first detailed  studies targeting high redshift galaxies with reasonably 
low assembled stellar masses (10$^{7}$ - 10$^9$ M$_\odot$) have begun to emerge 
(e.g., Erb et al. 2010, Brammer et al. 2012, Christensen et al. 2012, Maseda et al. 2013).   The rest-frame ultraviolet spectra show large 
equivalent width nebular emission lines which are rarely seen in spectra of more massive star forming systems 
(CIII]$\lambda\lambda$1907,1909\footnote{The CIII] doublet is actually a combination of [CIII]$\lambda$1907, a forbidden magnetic 
quadrupole transition and CIII]$\lambda$1909, a semi-forbidden electro-dipole transition.   
In  low resolution optical spectra of $z\simeq 2$ galaxies, the  doublet is  not resolved.   For the remainder of this paper, 
we will refer to the line as either 
the blended CIII]$\lambda$1908 doublet or simply as CIII]. }, OIII] $\lambda\lambda$1661, 1667, He II $\lambda$1640 and CIV $\lambda\lambda$
1548,1550), while metallicity measurements from rest-frame optical emission lines point toward metal poor ionised gas ($\lsim 0.2$ Z$_\odot$).  
The relative strengths of the emission lines require a much larger ionisation parameter than is seen in 
more massive, metal-rich galaxies.    It is not known what is driving the large ionisation parameter or what is causing 
the emission lines to be so prominent.   Moreover with the 
very small existing spectroscopic samples, it is unclear how common these ultraviolet emission features are in 
low mass high redshift galaxies.   

A potentially related development is the discovery of a substantial population of $z\simeq 1.5-2$ low mass 
star forming galaxies with very large (200-1000~\AA) rest-frame optical emission lines (e.g., Atek et al. 2011, van der Wel et al. 2011, 
Maseda et al. 2013, Masters et al. 2014, Amor{\'{\i}}n et al. 2014, Atek et al. 2014b).    
It has recently been argued that such extreme optical line emitting galaxies become ubiquitous among UV-selected 
star forming galaxies at $z\gsim 4$ (e.g., Shim et al. 2012, Stark et al. 2013a, Smit et al. 2013).   The 
equivalent widths of [OIII], H$\alpha$, and H$\beta$ likely require large specific star formation rates, 
as might be expected from dwarfs 
undergoing bursty activity.   Whether these strong optical line emitters are related to the low mass galaxies with 
extreme UV line emission lines described above is not yet known.   If there is a connection between the two populations however, 
then the presence of extreme optical line emission in $z\gsim 6$ galaxies (Smit et al. 2013) would indicate that the UV line spectra 
are likely to be much richer than anticipated.   Given the attenuation of Ly$\alpha$ in $z\gsim 6$ galaxies (e.g., 
Schenker et al. 2012, Ono et al. 2012, Pentericci et al. 2012, Treu et al. 2013, Schenker et al. 2014), the existence of prominent 
CIII] or OIII] emission might provide our best hope of characterising low mass $z\gsim 6$ galaxies with ground-based 
spectrographs. 

Clearly a much larger spectroscopic database of low mass galaxies is required to clarify the implications of 
the observations described above.  Fortunately with the large samples of faint gravitationally lensed galaxies that are now being discovered 
with deep imaging and grism spectroscopy from the {\it Hubble Space Telescope} (e.g., Richard et al. 2007, Limousin et al. 2007, Bradley et al. 2013, Schmidt et al. 2014, Atek et al. 2014a, Coe et al. 2014), it is becoming feasible to explore the  
properties of sizeable samples of low mass galaxies with unprecedented detail.  In Alavi et al. (2014),  a sample 
of ultra-faint gravitationally lensed galaxies in the Abell 1689 field was presented, demonstrating that the luminosity 
function rises steeply toward M$_{\rm{UV}}=-13$ at $z\simeq 2$.   Here we build on this progress 
by using Keck, VLT, and Magellan to obtain deep optical and near-infrared spectra for a subset of the  faint lensed galaxies in Abell 1689 and 
similarly faint samples in two other cluster fields (MACSJ0451+0006 and Abell 68).     

These spectra will allow us to determine whether the prominent UV emission lines seen in Erb et al. (2010) and 
Christensen et al. (2011) are common in low mass star forming galaxies at high redshift.    Through characterisation 
of the rest-optical emission lines, we will examine whether galaxies with large equivalent width UV emission 
lines have similar optical emission line spectra to the extreme line emitters reported in Atek et al. (2011) and 
van der Wel et al. (2011).    Using photoionisation models, we will attempt to understand what the powerful 
UV emission lines tell us about the nature of low mass galaxies at high redshift.     Our ultimate goals are twofold.
Firstly we seek to understand whether the spectra of low mass galaxies are consistent with the picture of  bursty   
star formation expected with strong stellar feedback.   And secondly, motivated by the very low success rate in recovering  
Ly$\alpha$ at $z\gsim 6$, we aim to determine whether UV lines such as 
CIII], CIV, and OIII] might be detectable in $z\gsim 6$ galaxy spectra.  

The paper is organised as follows.   In \S2, we describe the optical spectroscopic observations undertaken with 
Keck and the VLT and then discuss the sample selection, multi-wavelength imaging, magnification, and 
distribution of UV luminosities.    In \S3, we discuss the equivalent width distribution and flux ratios of UV emission lines of our 
sample of lensed galaxies.    With the goal of understanding what drives the UV emission line strengths, we 
characterize the stellar masses, UV continuum slopes,  metallicities, and relative chemical abundances of 
our sample in \S4.   In \S5, we use photoionisation models to investigate what range of properties (metallicity, age, 
ionisation parameter) are required to reproduce the UV emission line spectra of our low mass sample.   Finally in \S6 we 
assess implications for star formation in low mass galaxies at $z\simeq 2-3$ and discuss the feasibility and physical 
motivation for detecting lines other than Ly$\alpha$ with existing samples of $z\gsim 6$ galaxies.   We summarise our 
conclusions in \S7.

Throughout the paper, we adopt a $\Lambda$-dominated, flat universe
with $\Omega_{\Lambda}=0.7$, $\Omega_{M}=0.3$,  and
$\rm{H_{0}}=70\,\rm{h_{70}}~{\rm km\,s}^{-1}\,{\rm Mpc}^{-1}$.  We use a 
solar oxygen abundance of 12 + log (O/H) = 8.69 (Asplund et al. 2009).  All
magnitudes are quoted in the AB system (Oke et al. 1983).

\begin{table}
\begin{tabular}{llllc}
\hline  Cluster &  Instrument &  Setup  &  Dates & t$_{\rm{exp}}$    \\ 
 & & & & (ksec) \\ \hline 
MACS 0451  & FORS2 & G300V  & 27 Jan 2012 & 8.1 \\
MACS 0451  & FORS2 & G300V  & 28 Jan 2012 & 10.8 \\
Abell 68 & FORS2 & G300V & 20 Oct 2011 & 5.40  \\
Abell 68 & FORS2 & G300V & 15 Nov 2011 & 2.70  \\
Abell 68 & FORS2 & G300V & 18 Jul 2012 & 2.70  \\
Abell 68 & FORS2 & G300V & 19 Jul 2012 &  8.10 \\
Abell 1689 & FORS2& G300V  & 19 Feb 2012 & 2.70 \\
Abell 1689 & FORS2& G300V  & 20 Feb 2012 & 2.70 \\
Abell 1689 & FORS2& G300V  & 21 Feb 2012 & 2.70 \\
Abell 1689 & FORS2& G300V  & 15 Mar 2012 & 2.70 \\
Abell 1689 & FORS2& G300V  & 17 Mar 2012 & 2.70 \\
Abell 1689 & FORS2& G300V  & 18 Mar 2012 & 2.70 \\
Abell 1689 & FORS2& G300V  & 25 Mar 2012 & 2.70 \\
Abell 1689 & LRIS &  400/3400 (B) & 9 May 2010 &  18.0 \\
  \ldots &  \ldots & 600/7500 (R) & \ldots  &  12.6 \\
Abell 1689 & LRIS & 400/3400 (B) & 24 Feb 2012  &9.0   \\
 \ldots & \ldots & 600/7500 (R) &  \ldots   &  5.9\\
\hline
\end{tabular}
\caption{\label{obs}Details of optical spectroscopic observations. From left to right: cluster field, 
instrument used, instrument setup, observation dates, and total 
exposure time.  For the Keck/LRIS observations, the blue side set up is denoted as ``B" and red side as ``R."}
\end{table}

\section{Optical spectroscopy and Multi-Wavelength Imaging}

In this section we introduce an ongoing spectroscopic program targeting low luminosity 
gravitationally-lensed galaxies at high redshift.  We describe Keck observations in \S2.1 and VLT observations 
in \S2.2.   In \S2.3, we present the final spectroscopic sample.   We describe multi-wavelength imaging datasets in 
\S2.4, discuss the lensing magnifications and source luminosities in \S2.5, and present Magellan near-IR 
spectroscopy of a subset of our sample in \S2.6.

\subsection{Keck/LRIS}

Optical spectra of lensed galaxies in the Abell 1689 field were obtained with the the Low Resolution 
Imaging Spectrometer (LRIS; Oke et al. 1995) on the Keck I telescope on 09 May 
2010 and 24 February 2012.  Both nights were clear and the seeing was 0\farcs 7 (May 2010) and 1\farcs 0 (Feb 2012).  
The slit width was 1\farcs 2 for both masks. A dichroic was used to split the light 
at 5600 \AA\ between the two arms of the spectrograph.    On the blue side, a grism with 400 lines/mm blazed at 3400 \AA\ was used, and on the red side, 
the 600 l/mm grating blazed at 7500 \AA\ was used.  For the 1\farcs2 slit width used for both masks this results in a blue side resolution of 
8.2 \AA\  and a red side resolution of 5.6 \AA.   To reduce read
  noise, the blue side CCD was binned by a factor of 2 in the spatial direction. The blue and red side exposures were 1800 sec and 740 sec, respectively.  Total 
  integration times were 18 ksec (blue) and 12.6 ksec (red) for the May 2010 observations and 9.0 ksec (blue) and 5.9s ksec (red) for the Feb. 2012 observations. 
The individual exposures were rectified, cleaned of cosmic rays and stacked using the pipeline of Kelson (2003).  The one dimensional spectra were 
optimally extracted using the Horne (1986) algorithm.   Details are summarised in Table 1.

\subsection{VLT/FORS2}
The FORS2 spectrograph on the Very Large Telescope UT1 (Antu) was used to target lensed background 
galaxies in the field of three massive clusters: MACSJ0451+0006 (hereafter MACS 0451), Abell 68, and Abell 1689. 
The observations were performed as part of ESO program 088.A-0571 (PI: Richard), where we set up FORS2 in 
MXU mode to allow for more flexibility when targeting the most magnified sources near the cluster centre.   
We used the combination of the G300V grism and the 
order filter GG435 to cover the wavelength range 3800-9500 \AA\ with a spectral resolution of 10.3 \AA\, 
as measured from the FWHM of bright isolated sky lines.   Seeing ranged between 0\farcs8 and 0\farcs9 for 
all three masks.   A summary of basic observational details is provided in Table 1. 
The FORS 
reduction pipeline was used to perform flat-fielding, wavelength calibration, sky subtraction, and object extraction.   
Flux calibration was performed using standard A0V stars.

\subsection{Lensed galaxy sample}

\begin{table*}
\begin{tabular}{llcccccccc}
\hline   Name &  z$_{\rm{spec}}$ &  RA & DEC &  m$_{\rm{AB}}$ &  $\mu$  & M$_{\rm{UV}}$ &W$_{\rm{Ly\alpha},0}$ (\AA) &  W$_{\rm{CIII]},0}$ (\AA) & Redshift-ID  \\  \hline \hline
\multicolumn{10}{c}{MACS 0451} \\ \hline
 1.1 &  2.060 & 04:51:53.399 & +00:06:40.31 & 22.1  &$45.0\pm 2.5$ & $-18.6\pm0.1$&\ldots & $6.7\pm 0.6$ & IS abs, Opt em \\
 6.2 &  1.405 & 04:51:53.592 &  +00:06:24.96& 21.7 & $14.7\pm1.8$ & $-19.4\pm0.1$ &\ldots & $<1.2$ & IS abs, Opt em \\
 4.1 & 1.810 & 04:51:54.488 & +00:06:49.01 & 25.6 &$5.8\pm 1.2$ & $-17.1\pm0.2$&  \ldots  &  $10.0\pm 2.4$ & IS abs, Metal em  \\
3.1 & 1.904 & 04:51:55.438 &  +00:06:41.16 & 23.7 &$25.2\pm 4.9$& $-17.5\pm0.2$ &\ldots & $2.0\pm0.7$ & Opt em\\
\hline \multicolumn{10}{c}{Abell 68} \\ \hline
C4 & 2.622&00:37:07.657  &   +09:09:05.90 & 24.8 & $46\pm7$ &$-16.3\pm0.2$&  $36.6\pm5.7$ & $6.7\pm 2.1$ & Ly$\alpha$ em, Metal em \\ C20b & 2.689 &  00:37:05.405 & +09:09:59.14 &23.3 & $90\pm12$  & $-17.1\pm0.2$& $>$19.9 & $>$10.4  & Ly$\alpha$ em, Metal em\\  
\hline \multicolumn{10}{c}{Abell 1689} \\ \hline
 881\_329 & 1.559 & 13:11:31.543 & -01:19:45.88 & 25.9 & $75.1\pm12.4$& $-13.7\pm0.2$ & \ldots & $7.1\pm 3.1$ & Metal em \\
 899\_340 & 1.599 & 13:11:35.705 & -01:20:25.22 &23.9& $7.5\pm 1.3$& $-18.2\pm 0.2$ & \ldots & $5.1\pm1.4$ & Metal em, Opt. em  \\
 883\_357 &  1.702  & 13:11:31.882 & -01:21:26.10 & 23.3 &$13.0\pm 1.4$ & $-18.4\pm0.1$ & $76.0 \pm18.8$  & $6.5\pm0.7$  & Ly$\alpha$ em, Metal em \\
 860\_359 & 1.702 &13:11:26.426 & -01:21:31.22 & 24.5 &$4.3\pm 0.3$ & $-18.4\pm0.1$ & $163.8\pm25.5$ & $12.4\pm1.5$   & Ly$\alpha$ em, Metal em\\
 885\_354 & 1.705 & 13:11:32.405 & -01:21:15.98 & 23.2& $15.5\pm 3.0$& $-18.3\pm0.2$ &  $35.7\pm 4.6$ & $3.9\pm 0.6$& Ly$\alpha$ em, IS abs \\
 863\_348 & 1.834 & 13:11:27.350 & -01:20:54.82 & 24.1 & $35\pm 3.7$ & $-16.6\pm0.1$ & $73.1\pm8.6$  & $13.5\pm1.6$ & Ly$\alpha$ em, Metal em     \\ 
 876\_330 & 1.834 &  13:11:30.320 & -01:19:51.13 & 24.1& $27\pm 3.1$& $-16.9\pm0.2$ &$>$50.0 & $10.0\pm2.7$ & Ly$\alpha$ em, Metal em \\ 
 869\_328 & 2.543 & 13:11:28.690 & -01:19:42.69 & 23.3& $270\pm 61$& $-15.8\pm0.3$ & $4.3\pm0.4$ & $1.8\pm0.3$ & Ly$\alpha$ em, IS abs \\ 
  854\_344 & 2.663 & 13:11:24.982 & -01:20:41.57 & 23.4 & $6.2\pm0.3$& $-19.9\pm0.1$  & $45.3\pm 2.7$& $>$4.0 & Ly$\alpha$ em, IS abs \\  
 854\_362 & 2.731 &  13:11:24.982 & -01:21:43.53 & 24.3 & $2.8\pm 0.1$ & $-19.9\pm0.1$ & $129.6\pm 21.8$ &  $12.0\pm 3.2$ & Ly$\alpha$ em, CIII] em\\
 846\_340 & 2.976 & 13:11:23.141 &  -01:20:23.08 & 25.3 & $2.5\pm 0.1$& $-19.2\pm0.1$ & $86.4\pm24.8$ &$>$10.3 & Ly$\alpha$ em, break  \\
\hline
\end{tabular}
\caption{Properties of spectroscopic sample of lensed galaxies presented in this paper.   Each galaxy has a deep
Keck/LRIS (\S2.1) or VLT/FORS2 (\S2.2)  spectrum.    We report V$_{606}$-band magnitudes in MACS 0451 and 
R$_{702}$-band magnitudes in Abell 68.   In Abell 1689, we report i$_{775}$-band magnitudes for all sources where 
available. 
The lensing flux magnification ($\mu$) are derived from up-to-date cluster mass models 
(see \S2.5 for references).  
The rest-UV spectrum of 876\_330 was presented in Christensen et al. (2012a), where it was denoted 
Abell 1689 arc 31.1.  The primary means of redshift identification is listed in the right-most column.  'IS abs' refers to 
UV metal absorption lines,  'break' denotes detection of a Ly$\alpha$ continuum break, 'Metal em' refers to UV 
emission from metallic species (i.e., OIII]$\lambda$1661,1667, CIII]$\lambda$1908), and 'Opt em' denotes rest-optical 
emission lines (i.e., H$\alpha$, [OIII]$\lambda$5007).    Five of the galaxies listed above are multiple images.   These include 
881\_329 (image 36.1 in Limousin et al. 2007), 885\_354 (image 22.3), 863\_348 (image 12.2), 876\_330 (image 31.1), 
854\_344 (image 17.3). 
    }
\end{table*}

The Keck and VLT slit masks were filled with gravitationally lensed galaxies spanning a range of redshifts and magnitudes.  
We have targeted galaxies with and without spectroscopic redshifts.   Systems with known redshifts were 
identified through earlier campaigns targeting multiply-imaged systems (e.g., Santos et al. 2004, Broadhurst et al. 2005, 
Richard et al. 2007, Frye et al. 2007, Richard et al. 2014, in prep).   Redshift 
confirmation of the input spectroscopic sample was achieved through identification of Ly$\alpha$ emission or interstellar UV absorption lines.   
The remaining slits  were devoted to multiply-imaged galaxies without confirmed redshifts.  

We isolate a subset of these for analysis in this paper using three basic selection criteria.    
   First, to ensure emission line measurements are made at the correct rest-frame wavelengths, 
   we only consider galaxies with robust spectroscopic 
redshifts.    Most galaxies that fall in our `robust' category have their redshift confirmed from two separate 
spectral features (i.e., Ly$\alpha$ emission and metal absorption or continuum break).    For several systems, 
the spatial extent and asymmetry of Ly$\alpha$ prove sufficient 
for redshift confirmation.    Any galaxies on the Keck and VLT masks lacking a confident redshift 
determination are excised from the sample.   

Second, we require that CIII] falls in the spectral window covered 
by the Keck and VLT spectra.   This criterion restricts the redshift range to $1.3<z<4.0$.   
Objects outside this redshift range are not included in our sample.   We also excise objects with redshifts which place 
the primary emission lines of interest under strong atmospheric OH lines.   As the sky lines are strongest in the 
red side of the optical, this primarily impacts galaxies at $z\gsim 2.5$.   

And finally, to ensure reliable emission line equivalent width measurements, we only consider 
galaxies with UV continuum detections in the spectra.   Given the depth of the spectra, 
this requirement has the effect of limiting our sample to galaxies with apparent 
optical magnitudes brighter than $m\simeq 25-26$, placing a lower threshold on the 
UV flux of galaxies in our sample.   While there are several fainter objects on our masks with 
confirmed redshifts (from Ly$\alpha$), the faint continuum dictates that  additional 
emission lines can only be detected if equivalent widths are very large (i.e., $>$50~\AA\ rest-frame), 
well in excess of the equivalent widths expected for the UV metal lines.  Examination of the 
spectra of the faint Ly$\alpha$ emitters that we excise from our sample confirms the absence 
of additional emission lines at the 50~\AA\ level.

After removing galaxies that do not satisfy the three criteria listed above, we are left with a sample 
of 17  lensed galaxies.   As we will discuss  in \S3,  the rest-ultraviolet spectra of 
16 of the 17 galaxies reveal CIII] emission.    Colour images of the 16 CIII] emitters are shown in Figure 1. 
The redshifts range between $z=1.599$ and $z=2.976$.  The spectral features used for
 redshift identification are listed in Table 2.  The sample is 
comprised of four galaxies from the MACS 0451 field, two galaxies from the Abell 68 field, and 
eleven galaxies from Abell 1689 field.   While the MACS 0451 and Abell 68 observations are 
entirely based on VLT spectra, the spectroscopic sample toward Abell 1689 is made up of a 
mixture of Keck and VLT spectra.    Nine galaxies have deep Keck spectra, and five have deep VLT 
spectra.   For the three systems with both Keck and VLT observations, we examine both spectra, ensuring that 
the redshift identifications and emission line measurements are consistent.      
 \subsection{Multi-wavelength imaging}
\begin{figure*}
\begin{center}
\includegraphics[width=0.97\textwidth]{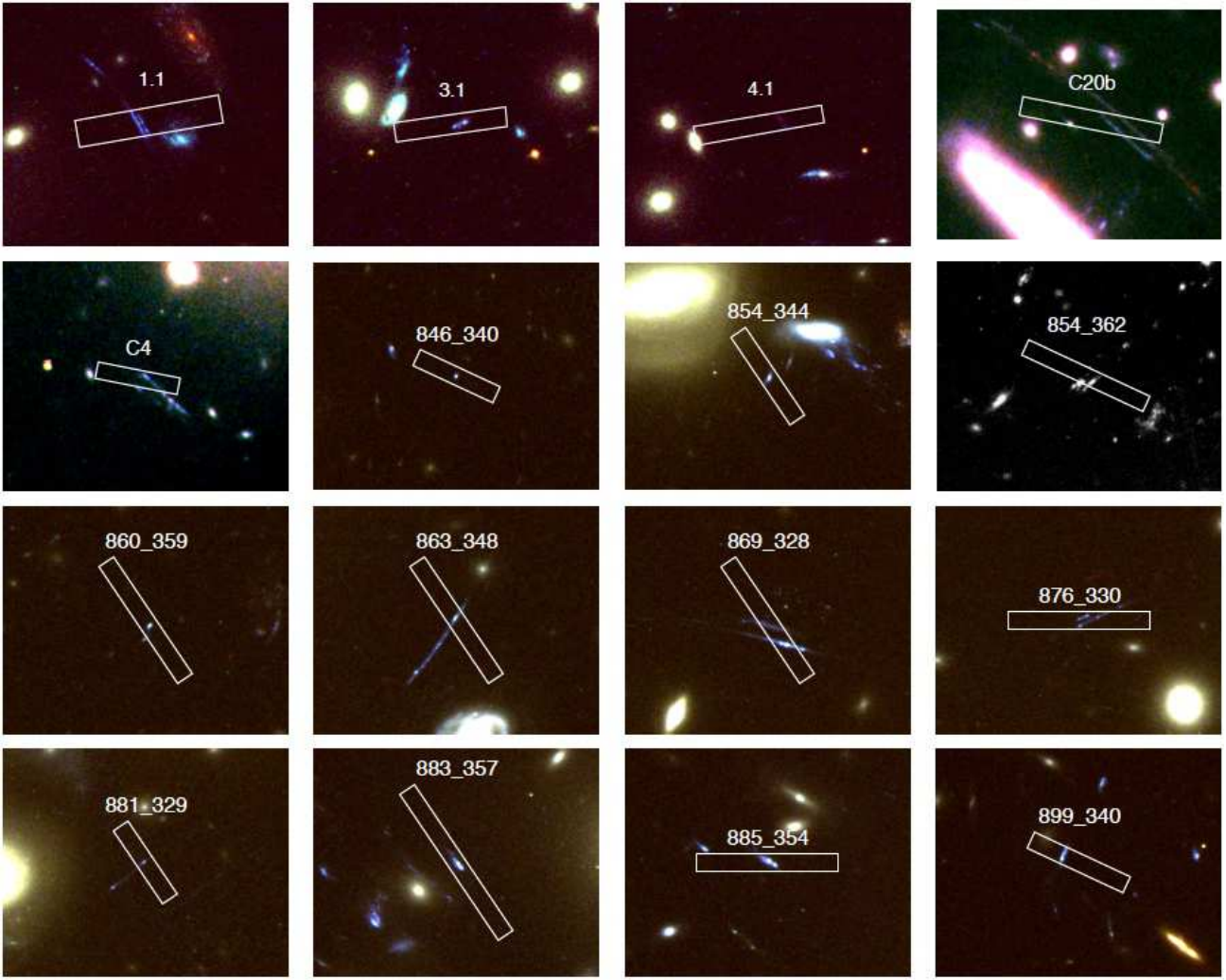}
\caption{Colour images of the 16 dwarf star forming galaxies at $1.6\lsim z\lsim 3.0$.   All of the galaxies have strong 
collisionally excited emission lines in the rest-frame ultraviolet.  We overlay the PA and slit positions of the 
Keck/LRIS and VLT/FORS spectra. }
\label{fig:montage}
\end{center}
\end{figure*}

Abell 1689 is the most well studied cluster field in our sample.  The optical portion of the SED is well sampled by  
deep HST/Advanced Camera for Surveys (ACS) imaging in four filters (F475W, F625W, F775W, F850 LP).
Details of the Abell 1689 ACS dataset have been provided in Broadhurst et al. (2005).    The near-infrared is constrained by a combination 
of HST/WFC3 imaging (F105W, F125W, F140W, F160W) and K$_s$-band imaging from the Infrared Spectrometer 
and Array Camera (ISAAC) formerly installed at the VLT.   For several sources, we also use constraints from 
HAWK-I photometry in the K band 
from VLT program 181.A-0185 (PI: Cuby).   The WFC3/IR and K$_s$-band imaging are described 
in detail in Richard et al. (2014, in preparation).   Deep {\it Spitzer}/IRAC data (total exposure time of 5 hours) have 
been obtained on Abell 1689 through the Spitzer program 20439 (PI:Egami).   We have used the 
publicly available Basic Calibrated from the Spitzer archive.

While not as well-studied as Abell 1689, the cluster MACS 0451 has been imaged in four filters by 
HST.    Deep ACS optical imaging has been taken in the F606W and F814W filters through programs 10491 and 12166 (PI: Ebeling).  
In the near-infrared, 
deep WFC3/IR imaging has been acquired in the F110W and F140W filters through program number 10875 (PI: Ebeling).    
Spitzer/IRAC imaging in the [3.6] and [4.5] 
bandpasses has been taken through IRAC cluster lensing program (ID 60034, PI: Egami).   Further details of the MACS 0451 imaging datasets 
will be provided in Richard et al. (2013, in prep).

Abell 68  has been observed with multiple optical and near-infrared filters with HST.  Deep R-band (F702W) imaging 
was obtained with the Wide Field Planetary Camera (WFPC2) during HST Cycle 8  (program 8249, 
PI: Kneib).   Details of the reduction of the F702W imaging are provided in Smith et al. (2005).    Additionally, 
Abell 68 has been imaged with ACS in the F814W filter, and in the near-infrared F110W and F160W filters 
with WFC3/IR as part of HST program 11591 (PI: Kneib). 

All HST data have been combined using the multidrizzle software (Koekomer et al. 2007) and aligned to the same 
astrometric reference frame.   For the {\it Spitzer} data, we have used the publicly available Basic Calibrated Data  
mosaicked onto a 0.6" pixel scale.

\begin{figure*}
\begin{center}
\includegraphics[width=0.49\textwidth]{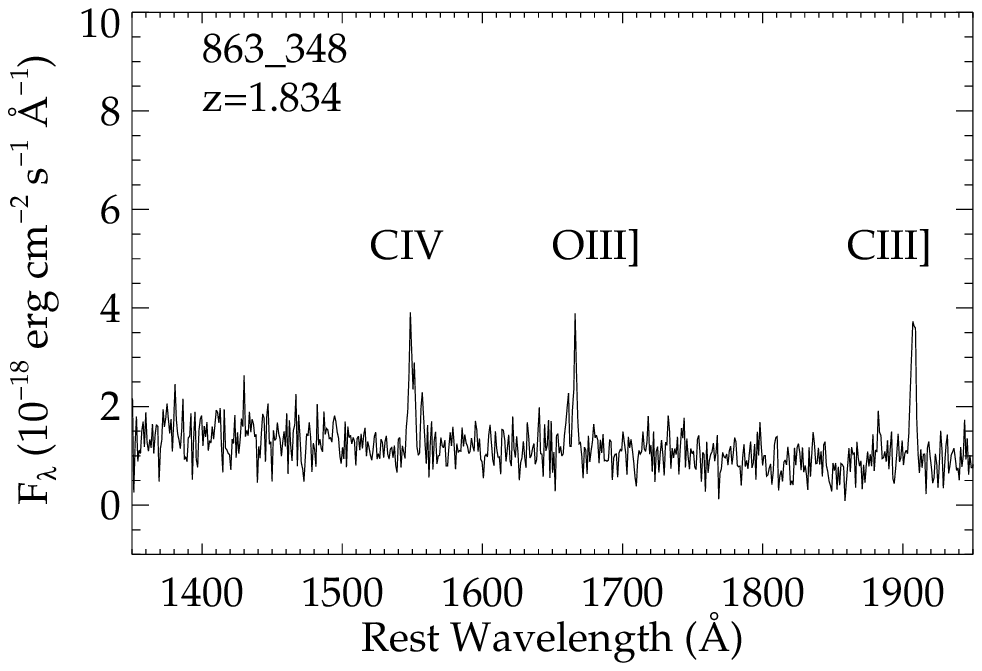}
\includegraphics[width=0.49\textwidth]{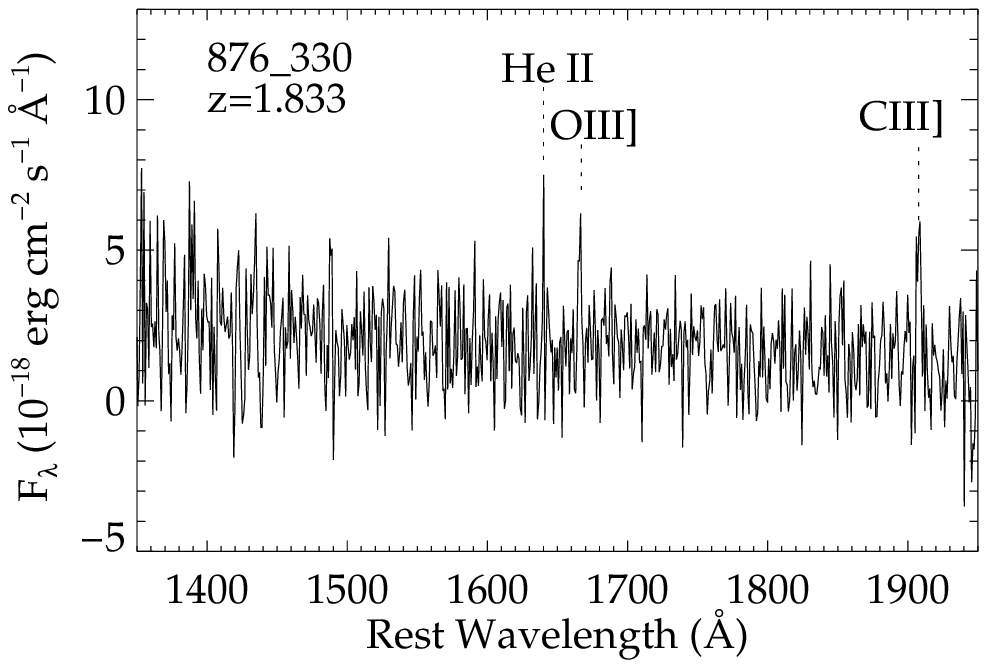}
\includegraphics[width=0.49\textwidth]{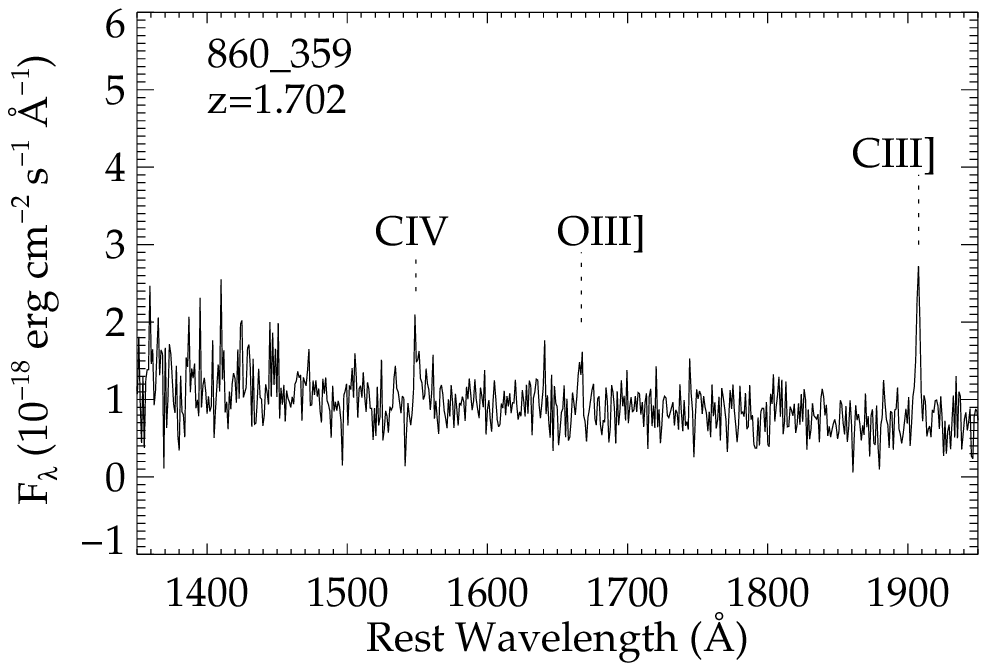}
\includegraphics[width=0.49\textwidth]{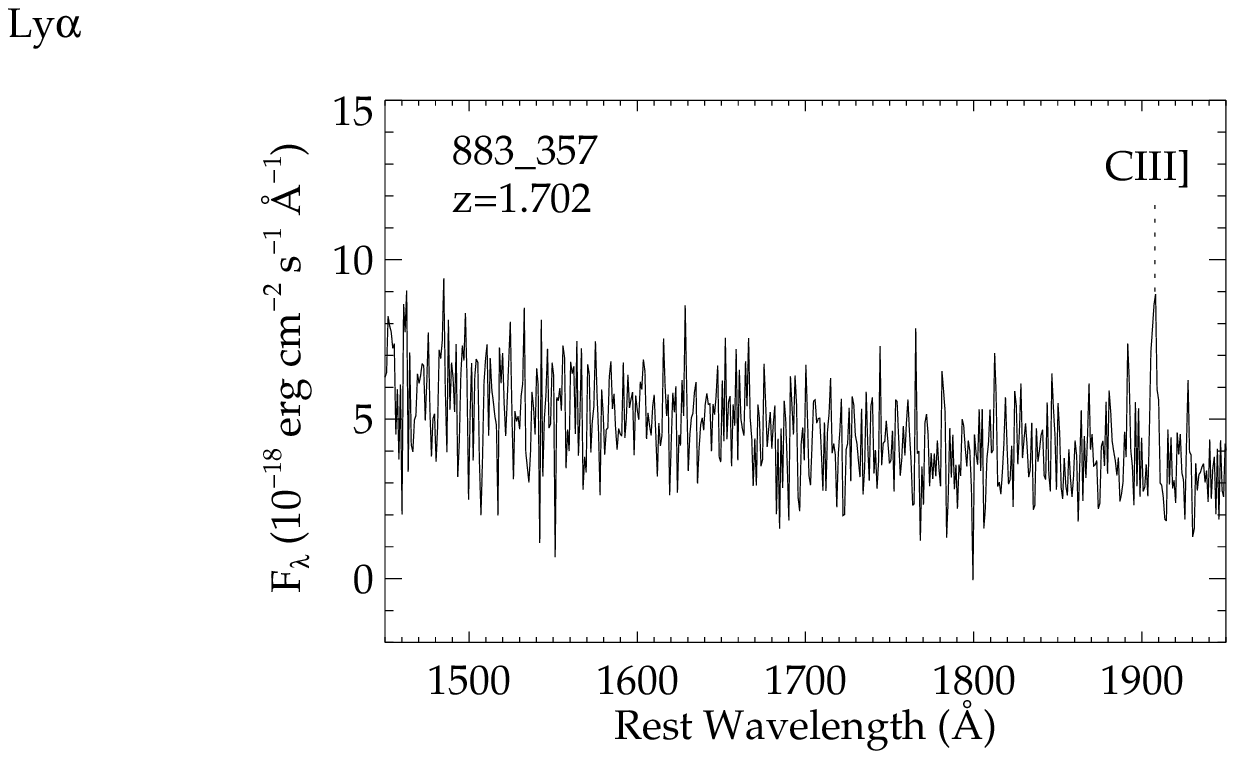}
\includegraphics[width=0.49\textwidth]{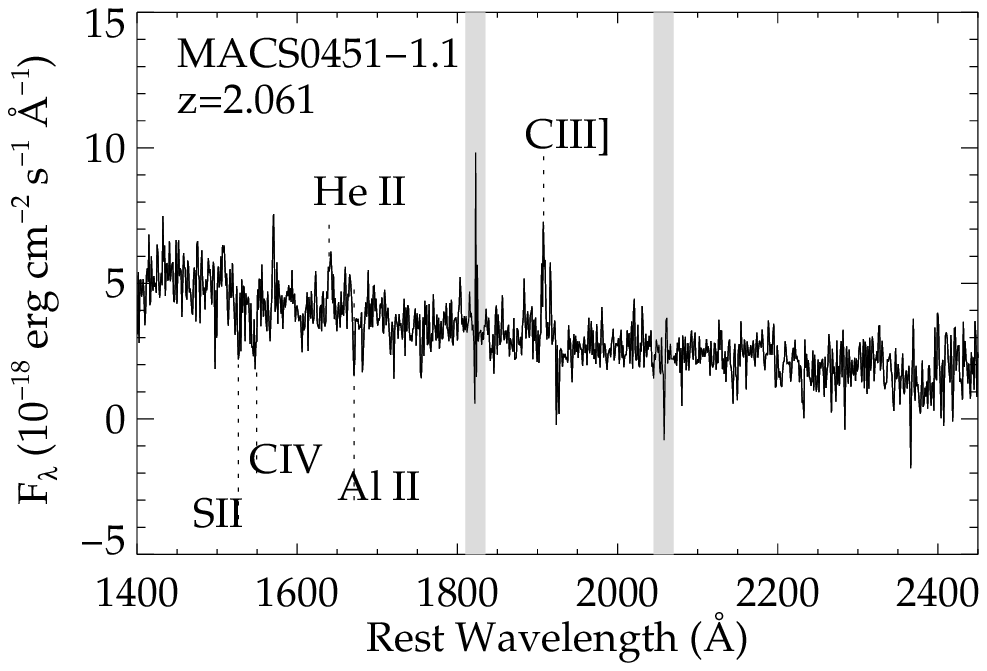}
\includegraphics[width=0.49\textwidth]{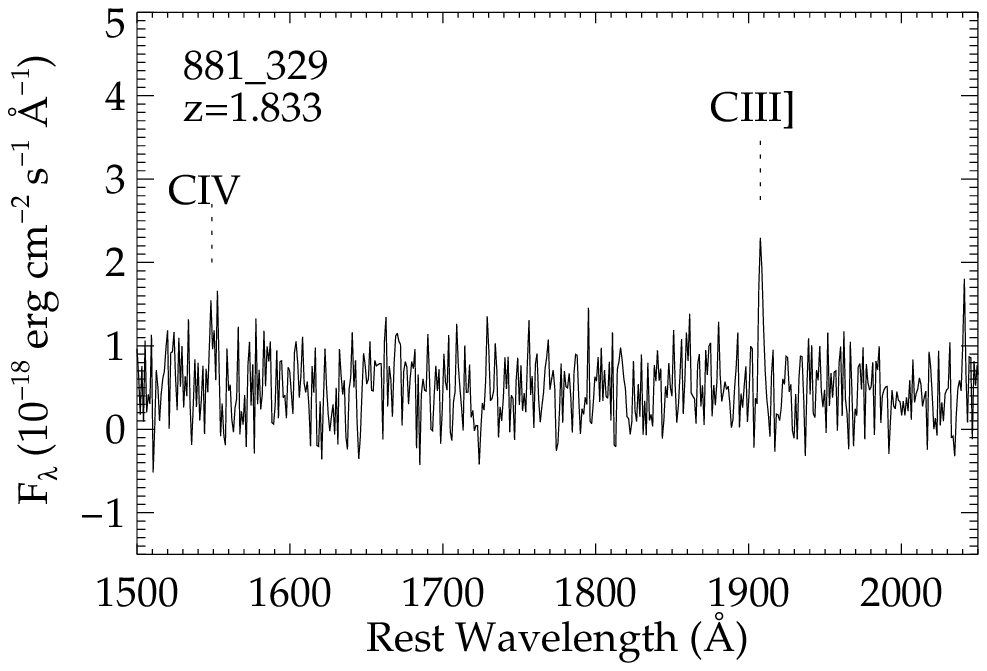}
\caption{Prominent emission lines in rest-UV spectra of intrinsically faint gravitationally-lensed galaxies.   The strongest 
line is typically the blended CIII]$\lambda$1908 doublet, but we often note emission from the blended CIV$\lambda$1549
doublet, He II$\lambda$1640, OIII]$\lambda\lambda$1661,1666, and Si III]$\lambda\lambda$1883,1892.  The fluxes are  
as observed, with no adjustment for lensing magnification.   Vertical grey swaths in MACS 0451-1.1 correspond to wavelengths 
with strong sky residuals.  
 }
\label{fig:spec1d}
\end{center}
\end{figure*}  
\subsection{Magnification and UV Luminosity}

Lensing magnifications ($\mu$) for each source are presented in Table 2.   The magnification factors 
and associated uncertainties are computed up-to-date cluster mass models using the LENSTOOL 
program (Kneib 1993, Jullo et al. 2007).   The cluster mass model used for Abell 1689 is based 
on Limousin et al. (2009) but is updated to include new spectroscopic redshifts of multiply-imaged galaxies.   
The mass model in Abell 68 is described in Richard et al. 
(2007) and Richard et al. (2010), while the MACS 0451 mass model is based on that presented in 
Jones et al. (2010), updated to include the  new multiple images presented in this work.   
Magnification factors range between $\mu=2.5$ and $\mu=270$.  The median magnification is $\mu=15.5$.   

The apparent optical magnitudes  (V$_{606}$  in MACS 0451, R$_{702}$  in Abell 68, and i$_{775}$ in Abell 1689) 
span m$_{\rm{AB}}$=21.7 and m$_{\rm{AB}}$=25.9 (see Table 2) with an 
average of m$_{\rm{AB}}$=23.9.   Without the magnification provided by lensing, this sample would have had an 
average apparent magnitude of m$_{\rm{AB}}$=26.9, which is much too faint for spectroscopic detection of low equivalent 
width emission features.  
The  absolute UV magnitudes of our sample range between M$\rm{_{UV}}$=$-13.7$ and M$\rm{_{UV}}$=$-19.9$ 
with a median of M$\rm{_{UV}}$=$-18.2$.  In addition to being considerably less luminous than most spectroscopic 
samples of unlensed UV-selected galaxies at $z\simeq 2-3$ (which generally span M$\rm{_{UV}}\simeq -20.0$ to $-22.5$), 
the galaxies in our sample are also intrinsically fainter than the large population of lensed systems that have been uncovered in 
the Sloan Digital Sky Survey (e.g., Diehl et al. 2009, Belokurov et al. 2009, Bayliss et al. 2011a, 2011b, Bayliss 2012, 
Stark et al. 2013b).    Deep spectra of the low luminosity galaxies in our sample thus provide a very complementary 
picture to existing samples of luminous lensed galaxies with deep rest-UV spectra (e.g., Pettini et al. 2002, 
Quider et al. 2009, Quider et al. 2010, Bayliss et al. 2013, James et al. 2013).   

\subsection{Near-Infrared Spectroscopy}

\begin{table}
\begin{tabular}{lclcc}
\hline  Cluster & Target &  Date & t$_{\rm{exp}}$ (ksec) & PA (deg)  \\  \hline 
MACS 0451 & 1.1a & 16 Feb 2012 &  2.7 & 125 \\
MACS 0451 & 1.1b  & 29 Oct 2012  & 3.0 & 30  \\
MACS 0451 & 6.2  & 16 Feb 2012 & 1.2 & 120    \\
MACS 0451 & 3.1  & 16 Feb 2012  &  0.6 & 120  \\
Abell 1689 &  899\_340 & 01 May 2013 & 3.6 & 345 \\
\hline
\end{tabular}
\caption{\label{obsnir}Summary of Magellan/FIRE near-infrared spectroscopic observations of lensed galaxies  
listed in Table 2.   Position angle of the slit is given in degrees E of N.    }
\end{table}

We have obtained near infrared spectra of four of the galaxies in Table 2 
using the Folded-port InfraRed Echellette (FIRE; 
Simcoe et al. 2010) on the 6.5 meter Magellan Baade Telescope.   An additional galaxy 
in our sample (860\_359) was 
observed for 1 hr with the XSHOOTER spectrograph (Vernet et al. 2011) on the VLT as part of ESO 
program 085.A-0909 (PI: Watson).   Details on the observational setup can be found in Amanullah et al. 
(2011).   

A summary of the FIRE observations 
is presented in Table 3.    Magellan/FIRE data were collected UT 16 Feb 2012, 29 Oct 2012, and 01 May 2013.  The typical seeing during 
the observations conducted on  2012 Feb 15 and 2012 Oct 28 was 0\farcs5 and 0\farcs8, respectively.   Conditions were variable 
on 01 May 2013 with seeing varying between 0\farcs6 and 1\farcs3.  
We used the echelle mode throughout both 
nights,  providing spectral coverage between
0.8 and 2.4~$\mu$m.   We used an 0\farcs 75 slit  on 16 Feb 2012 and 01 May 2013, delivering a resolving power 
of R=4800.   Skylines in the  February 2012 data are measured to have a Gaussian $\sigma$ of 0.9~\AA\  and 1.9~\AA\,  at 1.1 $\mu$m 
and 2.2 $\mu$m, respectively.   In the  May  2013 data, we measure $\sigma\simeq $1.3~\AA\ and 2.2~\AA\ at 1.1 $\mu$m 
and 2.2 $\mu$m.  The spectra  collected on 
28 Oct 2012 were obtained with a 1\farcs 0 slit width, providing a somewhat coarser spectral resolution ranging between 
$\sigma \simeq 1.5$~\AA\ and 2.5~\AA\ between 1.1 and 2.2~$\mu$m.

Reduction of the Magellan/FIRE spectra was performed using the FIREHOSE IDL pipeline developed for FIRE.   
Wavelength calibration was achieved using Th+Ar reference arc lamps.  For    
telluric absorption and relative flux calibration, we used spectral observations of A0V standard stars.  
One-dimensional flux and error spectra are extracted interactively for each object  using an aperture defined 
by the positions and widths of the rest-optical emission lines present in individual exposures. 

\section{Rest-frame ultraviolet spectra}
 
We show a subset of our highest quality optical spectra in Figure \ref{fig:spec1d}.   
 The most notable features of the spectra shown in Figure 1 are 
 the ubiquitous presence of ultraviolet emission lines (i.e., CIV, OIII], CIII]) and the near-absence of low and high ionisation interstellar 
absorption lines (i.e., Si II, CII, Si IV, CIV, Al II) which commonly appear in ultraviolet spectra.   
In this paper, most of our discussion will focus on the emission lines.   A detailed analysis of the 
absorption line properties will appear in a subsequent  paper.

\subsection{Characterisation of emission lines}

In each of the 17 galaxies in our sample, we search for emission from Ly$\alpha\lambda$1216, NV$\lambda$1240, 
NIV]$\lambda$1487, CIV$\lambda\lambda$1548,1550, He II$\lambda$1640, OIII]$\lambda\lambda$1661,1667,  NIII]$\lambda$1750, 
Si III]$\lambda\lambda$1883,1892, and the blended CIII]$\lambda$1908 doublet.   While no significant NV or NIII] emission is detected in 
any of our galaxies, each of the other lines is present in at least one of the galaxies in our sample.  

If an emission feature is detected significantly, line fluxes are determined through Gaussian fits to the observed 
emission.  For each emission line detection, the 1$\sigma$ error in the line flux  is 
determined through bootstrap resampling of the line within the allowed uncertainties.  
 In most cases, the continuum is detected with significance, allowing equivalent widths to 
be reliably determined.   In the few cases in which the continuum is detected at low significance ($S/N < 2$), we 
measure the 2$\sigma$ continuum flux density limit near the wavelength of the emission line in question, allowing 
us to place a lower limit on the emission line equivalent width.    Finally, in cases where the continuum is significantly 
detected but no flux from the emission line is detected, we note 2$\sigma$ upper limits on the flux and 
equivalent width.   The median 3$\sigma$ flux limit for CIII] is 4.3$\times$10$^{-18}$ erg cm$^{-2}$ s$^{-1}$.  In the highest quality 
spectra, the line flux limit is as low as 1.8$\times$10$^{-18}$ erg cm$^{-2}$ s$^{-1}$.    We measure CIII] fluxes ranging between 
2.2$\times$10$^{-18}$ erg cm$^{-2}$ s$^{-1}$ and 2.0$\times$10$^{-17}$ erg cm$^{-2}$ s$^{-1}$.

\subsection{CIII] EW distribution}

We find that the strongest rest-UV emission line (other than Ly$\alpha$) is always the blended CIII]$\lambda$1908 
doublet.  In 16 of our 17 galaxies, the CIII] emission line is significantly detected. 
The mean CIII] equivalent width of our sample (see Table 2 for individual values)  is 7.1~\AA,
similar to the value reported for the metal poor $z=2.3$ star forming galaxy in Erb et al. (2010).   We note that unlike 
the star formation rate or stellar mass, the equivalent width requires no adjustment for magnification.
In Figure 3, we present the distribution of CIII] equivalent widths (including results from BX 418 from 
Erb et al. 2010 and M2031 from Christensen et al. 2012a).      The composite UV spectrum of 811 $z\simeq 3$ LBGs in Shapley et al. (2003) 
provides a useful comparison.   This population is on average intrinsically brighter than the star forming dwarf galaxies considered 
in this paper.  While the Shapley et al. (2003) composite LBG spectrum yields a significant detection of CIII]$\lambda$1908, the equivalent width 
(1.67~\AA) is more than a factor of 4 lower than the average of the low luminosity star forming galaxies in Table 2.

But while low luminosity galaxies are certainly offset toward larger CIII] equivalent widths with respect to more luminous systems, it is 
clear from Figure 3 that there is a considerable dispersion in the equivalent width distribution.      On one hand, our sample includes 
a population of `extreme' CIII] emitters with equivalent widths  ranging between 10.0~\AA\ and 13.5~\AA.    But there are also  
low luminosity star forming systems in our sample with CIII] equivalent widths of just 1.8-3.9~\AA.     In the following sections, we will 
attempt to  understand both why CIII] equivalent widths appear to be larger in low luminosity star forming systems and  what physically is driving 
the large scatter in the CIII] equivalent width distribution.  

\begin{figure}
\begin{center}
\includegraphics[width=0.49\textwidth]{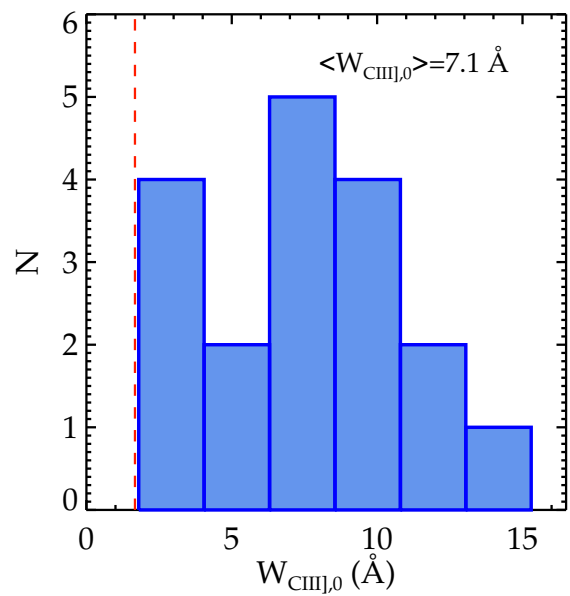}
\caption{Rest-frame CIII]$\lambda$1908 equivalent width distribution in low luminosity gravitationally lensed galaxies at $z\simeq 1.5-3.0$.   We include the 16 CIII] emitters in our sample and two similar systems (BX418 and M2031) reported in Erb et al. (2010) and Christensen et al. (2012a). The galaxies included here are 
metal poor with large specific star formation rates and blue UV slopes.     The red dashed line shows the CIII] equivalent width measured in the Shapley et al. (2003) 
composite spectrum of $z\simeq 3$ LBGs.   }
\label{fig:c3ew}
\end{center}
\end{figure}

\subsection{Additional rest-UV emission lines in CIII] emitters}

While CIII] is typically the strongest emission 
line other than Ly$\alpha$ in the rest-UV spectra of low luminosity galaxies at high redshift, the spectra shown in Figure 2 reveal 
additional emission features.   By characterising the flux ratios of the ultraviolet emission lines, we can gain additional insight into the 
ionised gas physical conditions and stellar populations of our sample of dwarf star forming galaxies.   In Table 4, we list the flux ratios 
relative to CIII] for galaxies in our sample with high quality spectra.  

Perhaps the most striking feature in the rest-UV spectra shown in Figure 2 is the presence of CIV $\lambda\lambda$1548,1550 emission, indicating that  
the  ionising spectra of several dwarf star forming galaxies in our sample have a substantial flux of photons with energies in excess of 47.9 eV, similar to 
what is often inferred from spectroscopy of nearby blue compact dwarf galaxies (e.g., Guseva et al. 2000, 
Thuan \& Izotov 2005, Shirazi \& Brinchmann 2012).
Accurate identification of CIV emission in high redshift star forming galaxies is often compromised in low resolution spectra 
by overlapping interstellar and P-Cygni absorption, but the absence of strong absorption features in many of our galaxies 
gives us a rare unobscured measure of the CIV strength.  The CIV/CIII] flux ratio is largest ($0.8\pm0.1$) in 
the $z=1.863$ galaxy 863\_348, the most extreme CIII] emitter in our sample.   Two other CIII] emitters in our 
sample show CIV detections with CIV/CIII] flux ratios of 0.4 (860\_359) and 0.5 (881\_329).  The rest-frame CIV 
equivalent widths of these three systems span 3.3-8.1~\AA.  

Ultimately we would like to characterise the distribution of nebular CIV/CIII] flux ratios (an indicator of the hardness of the 
ionising spectrum) in dwarf star forming galaxies.    While CIV is only significantly detected in 3 of 17 systems, it should be emphasized that 
non-detection of CIV  in our sample does not necessarily indicate a low ($\lsim$0.5) CIV/CIII] ratio.   Given that CIV 
appears to typically be at least a factor of two fainter than CIII], we can only detect  CIV emission in those galaxies with very large equivalent width 
CIII] emission.  In the eight galaxies with rest-frame CIII] equivalent widths in excess of 7.0~\AA, three systems have 
CIV detections with CIV/CIII] flux ratios greater than 0.4 (881\_329,860\_359,863\_348).   One of the undetected galaxies 
has a strong CIV absorption component (854\_362) which precludes identification of nebular emission in our low resolution spectra.   
In the remaining 4, we place 2$\sigma$ upper limits of 0.6-2.3 on the CIV/CIII] flux ratios and thus cannot rule out the presence of 
CIV at 40-50\% the flux of CIII].    Thus among the large equivalent width CIII] emitters, powerful CIV emission appears to be present with a 
CIV/CIII] flux ratio greater than 0.4 at least 50\% of the time.

Given that the energy required to doubly ionise helium  (54.4 eV) is only slightly greater than the energy required 
to triply-ionise carbon (47.9 eV), we might expect to detect nebular He II$\lambda$1640 
emission in several of the extreme CIV emitting galaxies in our sample.   But as is clear from Figure 2, many of the extreme CIII] and 
CIV emitting galaxies in our sample do not have detectable He II emission.   The best limit comes from the lack of strong He II emission in 863\_348, 
the galaxy with the largest equivalent width CIII] and CIV emission.   Adopting a  
2$\sigma$ upper limit on the He II flux, we find that 863\_348 has a He II/CIII] ratio of $\lsim 0.1$ and a rest-frame equivalent width of 
$\lsim 1.4$~\AA.    The only confident He II detection is in the CIII] emitting galaxy MACS 0451-1.1.   The He II / CIII] flux ratio in this system 
($0.5\pm0.1$) is very similar to that found for the metal poor galaxy BX 418 in Erb et al. (2010).    As with BX 418, stellar winds might 
contribute significantly to the He II line in MACS 0451-1.1.   A higher resolution (and larger S/N) spectra would be required to 
disentangle the nebular and stellar components of the line.   

After CIII], the most commonly detected emission line (other than Ly$\alpha$) is the OIII]$\lambda\lambda$1661,
1666 doublet.  We commonly detect OIII]$\lambda$1666 emission with fluxes as large as 50-60\% those of CIII].   The OIII]$\lambda$1661 
emission line is generally weaker than OIII]$\lambda$1666, with fluxes typically not greater than 20\% those of CIII].    Among the 
five systems with the best quality spectra (and those listed in Table 4), 
rest-frame equivalent widths range between $\lsim 0.5$~\AA\ and 1.5~\AA\  for OIII]$\lambda$1661 and 1.2~\AA\  and 7.4~\AA\  
for OIII]$\lambda$1666.  The OIII]$\lambda$1666 equivalent widths in our sample are considerably greater than the equivalent 
width (0.23~\AA) measured in luminous star forming galaxies at $z\simeq 3$ (Shapley et al. 2003).   The most extreme UV line 
emitters (863\_348, 876\_330), show OIII]$\lambda$1666 emission with equivalent widths between 3 and 6$\times$ greater 
than that measured in the metal poor star forming system BX 418 (Erb et al. 2010).  

We also note the presence of weak emission from NIV] $\lambda$1487 and Si III] $\lambda\lambda$1883,1892 in 
several of our sources.  Where detected, the Si III] doublet typically is seen with a flux of 10-20\% that of CIII].   
Recent work has revealed strong NIV] emission in a strong Ly$\alpha$ emitting galaxy at $z=5.6$ (Vanzella et al. 2010) 
and in a powerful emission line target at $z=3.6$ (Fosbury et al. 2003).   While we do detect NIV] in 876\_330 
(with a flux $0.5\pm 0.2$ that of CIII]), we do not detect it in any of the other extreme line emitters.

Finally, many of the emission features seen in our low luminosity sample are also often observed in narrow-lined AGNs.     
The AGN composite spectrum constructed in Hainline et al. (2012) suggests an average CIV equivalent width of 
16.3~\AA, more than a factor of two larger than the most extreme sources in our sample.   The narrow-line AGN 
composite also shows NV (5.6~\AA) and He II (8.1~\AA), both of which are not seen in most of the low luminosity galaxies 
in our sample.  An additional point of comparison is the CIV/CIII] ratio which probes the hardness of the radiation field.   
The average CIV/CIII] flux ratio found for narrow-line AGN in the `class A' composite spectrum 
presented in Alexandroff et al. (2013) is 7.5, more than an order of magnitude larger than most of the galaxies in our sample.   
Thus while the low luminosity systems studied in this paper may have a harder spectrum than more 
luminous galaxies at high redshift, they are considerably less extreme than  typical narrow-lined AGNs.    However we 
do note that there are several narrow line AGN in the sample considered in Alexandroff et al. (2013) with CIV/CIII] ratios close to those 
spanned by our sample (see their Figure 8), so AGN contribution cannot be discounted.   In 
\S5, we will explore whether the high ionisation emission features seen in Figure 2 can be powered entirely by 
low metallicity stars without requiring radiation from additional energetic sources (i.e. fast radiative shocks, X-ray binaries, AGN).    Ultimately 
more data will be required to constrain the possible contribution of AGN.

\begin{table}
\begin{tabular}{lccccc}
\hline \hline  &  A1689 &  A1689  & A1689  &    M0451 & A1689\\
\hline  f$_{\rm{line}}$/f$_{\rm{CIII]}}$ &  876\_330&  863\_348&  860\_359 &   1.1 & 881\_329 \\  \hline
NV$\lambda$1240  &   $\lsim 1.1$ & $\lsim 0.2$   &  $\lsim 0.6$ &  \ldots & $\lsim 2.9$ \\
NIV]$\lambda$1487  & 0.5$\pm$0.2  & $\lsim 0.1$  & $\lsim 0.1$  &  \ldots  &  $\lsim 0.7$ \\
CIV$\lambda$1549 & $\lsim$ 0.6 & 0.8$\pm$0.1 & 0.4$\pm$0.1 &  \ldots & 0.5$\pm$0.3 \\
He II$\lambda$1640 & $\lsim$ 0.5 & $\lsim 0.1$ & $\lsim 0.2$ & 0.5$\pm$0.1& $\lsim 0.5$\\     
OIII]$\lambda$1661  & $\lsim$ 0.4 & 0.2$\pm$0.1 & $\lsim 0.2$ &0.2$\pm$0.1 &$\lsim 0.5$ \\
OIII]$\lambda$1666  & 0.5$\pm$0.2 & 0.6$\pm$0.1 & 0.3$\pm$0.1  &0.3$\pm$0.1 & $\lsim 0.5$ \\
NIII]$\lambda$1750  &   $\lsim 0.3$  & $\lsim 0.1$   &   $\lsim 0.1$ &  $\lsim 0.1$ & $\lsim 0.3$  \\
$\rm{[Si III]\lambda}$1883  & $\lsim$ 0.3&  0.2$\pm$0.1 & 0.1$\pm$0.1 &  0.2$\pm$0.1 &$\lsim 0.5$\\
$\rm{[Si III]\lambda}$1892 &  $\lsim$ 0.3  &  $\lsim$ 0.1  & 0.1$\pm$0.1  &0.2$\pm$0.1 &$\lsim 0.5$ \\
\hline
\end{tabular}
\caption{Rest-UV emission flux ratios (relative to the blended CIII]$\lambda$1908 doublet).    
If line is not detected significantly, we list the 2$\sigma$ upper flux ratio limit.   No CIV$\lambda$1549 
flux measurement is possible for MACS0451-1.1 owing to prominent interstellar absorption 
feature.    }
\end{table}

\subsection{Ly$\alpha$ emission of CIII] Emitters}

The CIII] emitters in our sample are almost always accompanied by powerful Ly$\alpha$ emission (Figure 4a) with 
rest-frame equivalent widths  ranging between 50 and 150~\AA\ (see Table 2).   The link between the 
equivalent width of CIII] and that of Ly$\alpha$ was first realised through examination of $z\simeq 3$ 
LBG spectra in Shapley et al. (2003).   By grouping galaxies by their Ly$\alpha$ equivalent width, 
Shapley et al. (2003) demonstrated  that the CIII] equivalent width increases from 0.41~\AA\ to 5.37~\AA\ 
as the Ly$\alpha$ equivalent width increases from $-14.92$~\AA\ (absorption) to 52.63~\AA\ (emission).   

The 17 low luminosity galaxies considered in this paper extend the results of Shapley et al. (2003) to larger Ly$\alpha$ and 
CIII] equivalent widths (Figure 4b), confirming the close relationship between the two quantities.  Our spectra generally 
reveal Ly$\alpha$ equivalent widths in excess of 50~\AA\  for galaxies with CIII] emission equivalent 
widths greater than 5~\AA, similar to the largest Ly$\alpha$ equivalent width bin in Shapley et al. (2003).  
 The most extreme CIII] emitters in our sample (with equivalent widths of 10-14~\AA) 
typically show Ly$\alpha$ equivalent widths of 75-150~\AA.  Notably, the galaxies in our sample with the lowest equivalent 
width CIII] emission ($\lsim 3$~\AA) also have low equivalent Ly$\alpha$ emission.      An exception to this relationship has recently been presented in Bayliss et al. (2013).   The $z=3.6$ galaxy presented in that paper shows numerous UV emission lines but 
Ly$\alpha$ is seen in absorption.    The system is considerably more massive (log M$^\star$ = 9.5) than galaxies in our sample.    
It is possible that large covering fractions of neutral hydrogen (and hence spectra with Ly$\alpha$ in absorption) 
might be more common among the more massive UV line emitters.

We compute the typical velocity offset between Ly$\alpha$ and CIII] assuming a mean blended CIII] rest-frame 
wavelength of 1907.709~\AA.   While our results do demonstrate that Ly$\alpha$ is typically redshifted 
with respect to CIII], the velocity offsets 
(60-450 km s$^{-1}$ with a mean of 320 km s$^{-1}$) are slightly lower than the average value ($\sim 450$ km s$^{-1}$) for 
UV continuum selected galaxies at 
$z\simeq 2$ (Steidel et al. 2010) but are comparable to Ly$\alpha$-selected galaxies (e.g., Taken et al. 2007, Hashimoto et al. 2013).   
We will come back to the importance of the velocity offset in \S6.1, when 
assessing the case for targeting CIII] in $z\gsim 6$ galaxies. 

The  tight relationship between the equivalent width of Ly$\alpha$  and 
CIII]  suggests that conditions which support the efficient production and escape of Ly$\alpha$ radiation (i.e., low dust content, 
low metallicity, partial coverage of neutral hydrogen) are either directly or indirectly linked to the production of collisionally-excited 
emission lines such as CIII].      With observations indicating that Ly$\alpha$ becomes more common in galaxies at 
earlier times over the redshift range $3\lsim z\lsim 6$ (Stark et al. 2010, Stark et al. 2011), we may   
expect large equivalent width CIII] emission to become ubiquitous among the galaxy population at $z\gsim 6$.   

\begin{figure*}
\begin{center}
\includegraphics[width=0.49\textwidth]{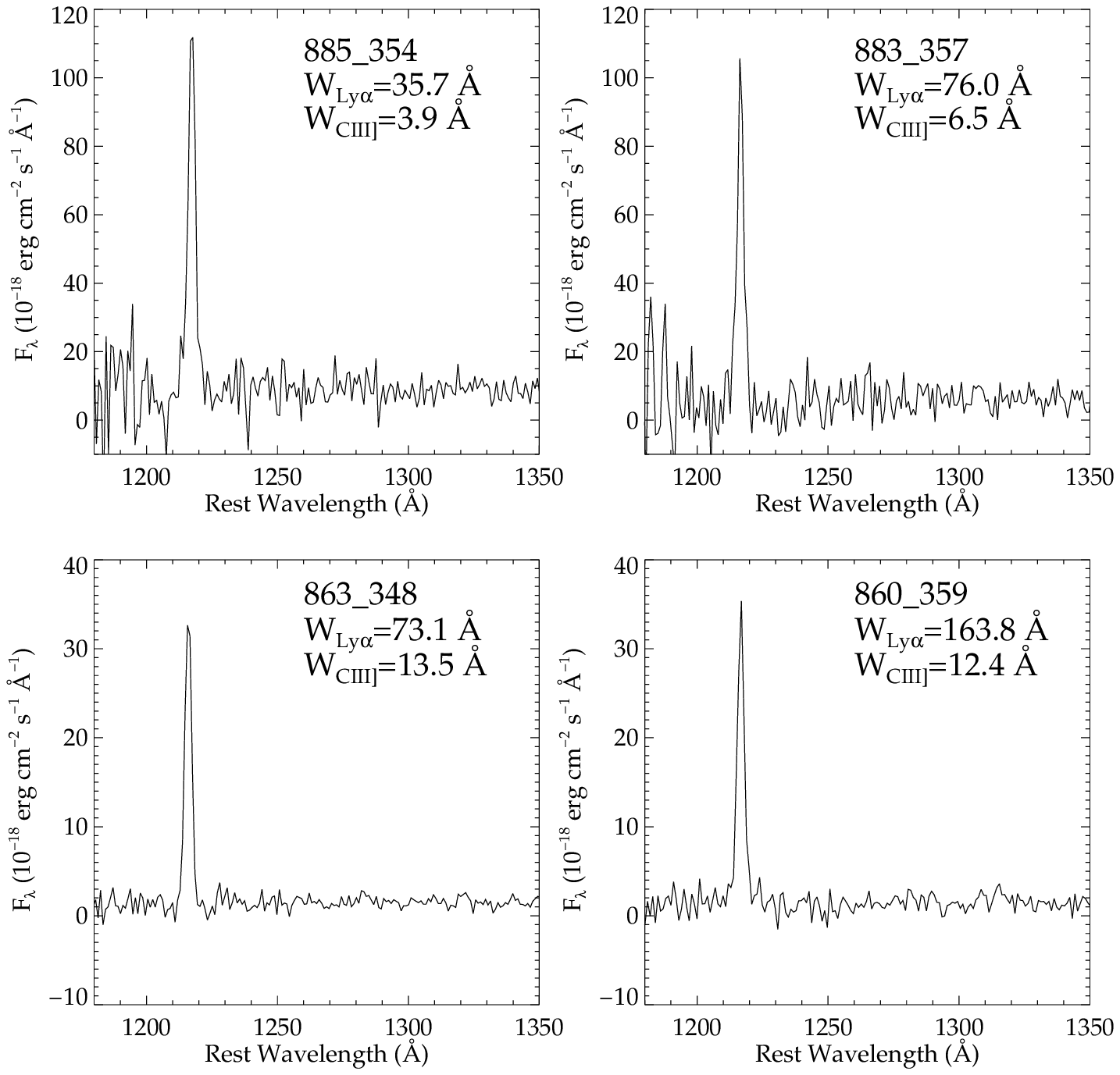}
\includegraphics[width=0.49\textwidth]{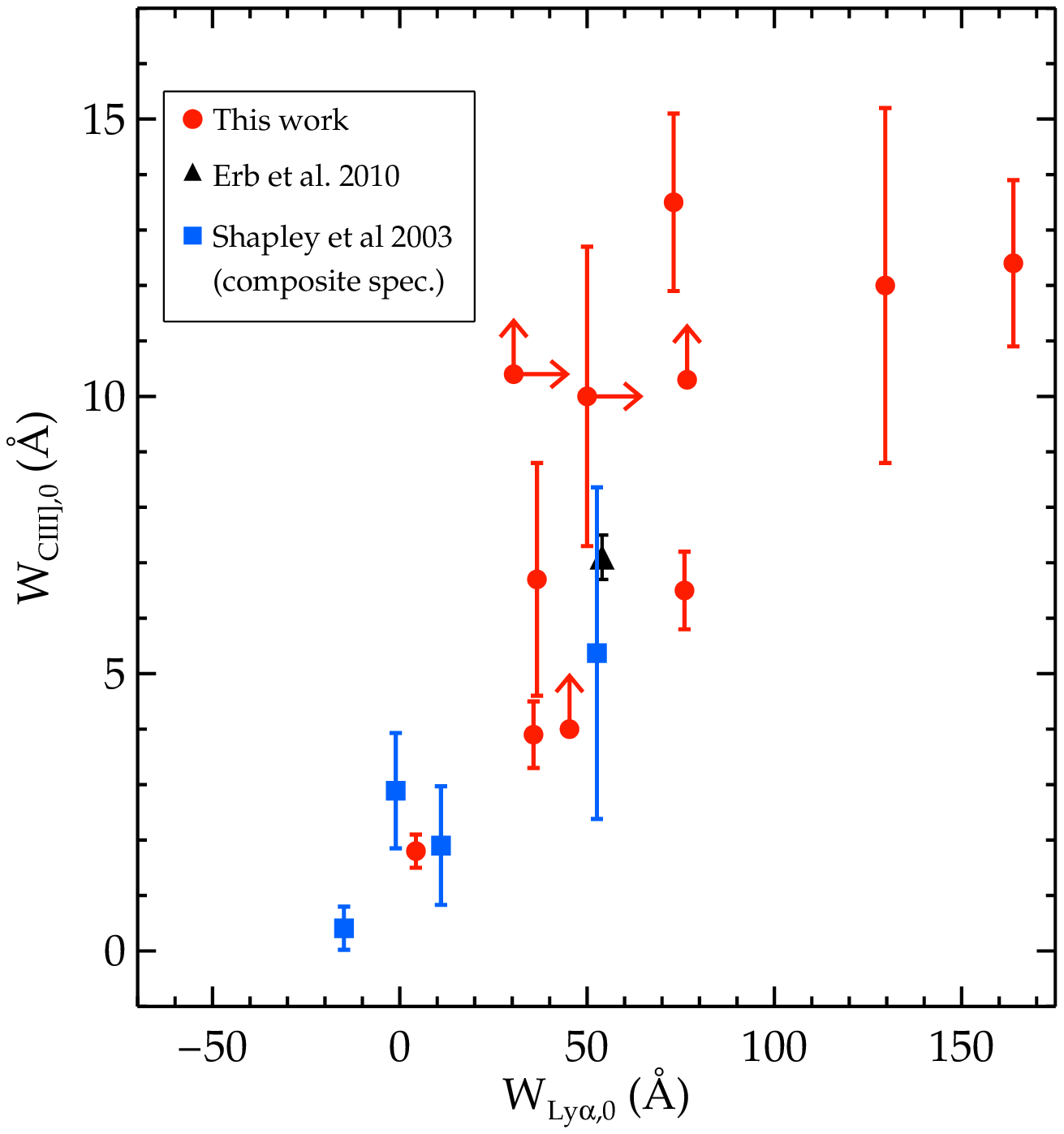}
\caption{Ly$\alpha$ emission in lensed galaxies with large equivalent width CIII] emission.  ({\it Left}:)  
Spectra of extreme CIII] emitting galaxies at rest-frame wavelengths 1180-1350~\AA.   
Ly$\alpha$ rest-frame equivalent widths are near the maximum expected values (50-150~\AA) 
for normal stellar populations.   No sign of NV$\lambda$1240 
emission is present.   Underlying continuum emission is detected with significance, but low 
ionisation absorption lines (Si II$\lambda$1260, OI+Si II$\lambda$1303, CII$\lambda$1334) are 
not detected.  ({\it Right}:) Relationship between 
CIII]$\lambda$1908 and Ly$\alpha$ equivalent width.  The low mass lensed galaxies in our sample with Ly$\alpha$ 
and CIII] measurements are denoted with red circles.  We also overplot measurements for BX 418 (Erb et a. 2010) 
as a black triangle and for the $z\simeq 3$ composite LBG spectra (Shapley et al. 2003) as blue squares.     }
\label{fig:lya}
\end{center}
\end{figure*}

\section{Physical Properties of CIII] emitters}

In this section, we seek to improve our understanding of what causes ultraviolet emission lines to have such 
large equivalent widths in low luminosity galaxies.   Based on the low luminosities, we expect the 
ionised gas to be metal poor (see \S4.5), 
elevating the electron temperature and boosting the strength of collisionally excited emission lines (OIII], CIII]).   But 
other aspects of the stellar populations and gas physical conditions may contribute significantly as well.   
In this section, we  investigate the physical properties (stellar mass, specific star formation rate, metallicity, 
carbon-to-oxygen abundance) 
of our population of strong CIII]-emitting galaxies with the goal of building a more complete picture of
 what might be driving the strong ultraviolet emission lines.   We also investigate whether the  
sample  of dwarf galaxies with powerful ultraviolet line emission is similar to the growing population of 
extreme optical line emitting galaxies.

\subsection{Population synthesis modeling}

We infer the stellar mass and specific star formation rate of individual systems through comparison of the observed 
SEDs with population synthesis models.  The procedure we adopt is similar to our earlier work (Stark et al. 2009, 
2013a).   We consider models with only stellar continuum and those that additionally include the contribution from  
nebular emission lines and continuum.   The nebular line contribution at each age step is calculated from a code 
presented in Robertson et al. (2010).   In this code, the hydrogen line intensities are computed from the case B 
recombination values tabulated in Osterbrock \& Ferland (2006), while the intensities of metal lines are computed using 
empirical results of Anders et al. (2003).   The stellar continuum is computed from the models of Bruzual \& Charlot (2003).    
For further details on the modelling procedure, see the description in Roberston et al. (2010) and Stark et al. (2013a).  

For consistency with modelling of high redshift star forming systems with larger UV luminosities (e.g., Reddy et al. 2012), 
we infer properties assuming a constant star formation history.    If star formation histories of the galaxies in our sample 
are bursty and episodic, as motivated by many simulations of dwarf galaxies (e.g., Teyssier et al. 2013, 
Hopkins et al. 2013, Shen et al. 2013) and discussed in recent work on EELGs (e.g., Atek et al. 2014c), our simple constant star formation history modelling will be in error.   
Neglecting the presence of an older stellar component from an earlier episode of star formation can 
cause the stellar mass to be underestimated.   When two component star formation histories including a young 
episode of star formation on top of an underlying very old stellar population are used in the SED fitting procedure, 
the inferred stellar masses of star forming $z\simeq 2-3$ galaxies can be 3$\times$ larger than derived using single component
star formation histories (e.g., Papovich et al. 2001, Shapley et al. 2005, Dominguez et al. 2014, in prep).     In the following, we will 
report physical properties derived assuming single component star formation histories (as these provide a satisfactory 
statistical fit to the continuum SEDs).   As we acquire more measurements of rest-optical emission lines in our sample, we will explore whether 
simultaneous fitting of the observed continuum and emission lines (e.g., Pacifici et al. 2012) provides useful 
constraints on how much the current episode of star formation contributes to the total stellar mass.

We adopt a stellar metallicity of 0.2 Z$_\odot$ for our grid (consistent with the gas-phase measurements we 
provide in \S4.5)  and a Salpeter (1955) initial mass function (IMF) spanning 0.1-100 M$_\odot$.  
Adopting a Chabrier IMF would  reduce the star formation rates and stellar masses by 1.8$\times$.     Allowing lower 
metallicities in the grid would not strongly affect the stellar masses (e.g., Papovich et al. 2001).   Dust reddening 
of the stellar continuum is included following Calzetti et al.  (2000).  Our model grid contains SEDs with differential 
extinction ranging between E(B-V)$\rm{_{\rm{stars}}}$=0.00 and E(B-V)$\rm{_{\rm{stars}}}$=0.50 in steps of $\Delta$E(B-V)=0.01.    
Recent work has confirmed that the extinction toward HII regions is greater than that of the stellar continuum in star forming galaxies 
at high redshift (e.g., Forster Schreiber et al. 2009, Wuyts et al. 2011, Price et al. 2013), similar to measurements of star forming 
galaxies in the local universe (Calzetti et al. 2000).    But the extra attenuation faced by the HII regions 
appears to decrease with increasing sSFR (e.g.,  Wild et al. 2011, Price et al. 2013).    The stellar continuum and 
nebular line emission in galaxies with very large sSFR ($\gsim 2.5$ Gyr$^{-1}$) are found to have similar levels of attenuation at high redshift 
(Price et al. 2013).    As we will show in \S4.2, the continuum SEDs of  galaxies in our sample imply very large sSFRs independent 
of assumptions about dust.   Motivated by the results of Price et al. (2013), we thus assume the nebular line and 
continuum attenuation is the same as the stellar continuum.  

In computing the stellar masses and star formation rates, 
we consider models with ages spanning between 30 Myr and the age of the universe at the redshift of the galaxy we are modelling.   
This lower age bound is often adopted based on dynamical timescale arguments, but at lower masses, it is 
conceivable that younger age models might be reasonable.  Reducing the minimum age (to i.e., 1-5 Myr) would  
introduce models with bluer intrinsic UV slopes.  As a result, more reddening (and hence more star formation) would be 
required to fit a given observed SED.   Furthermore, at very young ages for constant star formation histories, 
the ratio of UV luminosity and star formation rate has yet to equilibrate.   Because of this, models with very young ages 
generally require much more star formation to reproduce an observed UV flux density.    In the following, we 
will take the conservative approach of adopting a 30 Myr lower age bound, but we will also quantify the extent to which 
younger  models would alter the inferred properties. 

For each object in Table 2, we compare the galaxy SED to the model grid described above and  
compute the age, dust reddening, and normalisation which provides the best 
fit to the observed SED.    Using the normalisation and lensing magnification factors, we derive the star formation 
rate and stellar mass from the mass to light ratio of the best-fitting model template.   Uncertainties in the mass and 
star formation rate are derived by bootstrap resampling the observed SED within the allowed photometric errors.

\begin{table}
\begin{tabular}{llclc}
\hline  ID &log (M$_\star$/M$_\odot$) & $\beta$ &      sSFR & W$\rm{_{[OIII]+H\beta}}$  \\  
& &  & (Gyr$^{-1}$)  & (\AA) \\
\hline  \hline
\multicolumn{5}{c}{MACS 0451} \\ \hline
 1.1 & 7.69(7.80) & -2.7 & 28(35) & \ldots  \\
 6.2 &  9.00(9.12)  & -1.2  & 23(35) & \ldots  \\
4.1 &   7.51(7.58) & -1.5 & 34(34) & \ldots \\
 3.1 &  7.77(8.09) & -2.2 & 27(4.1)& \ldots \\ \hline
\multicolumn{5}{c}{Abell 68} \\ \hline
 C4  & 7.79(7.84) & -2.7 & 18(19) & \ldots \\
 C20b &  6.63(6.69) & -2.3 & 23(20) & \ldots \\ \hline 
\multicolumn{5}{c}{Abell 1689} \\ \hline
881\_329 & 6.30(6.54) & -2.5 & 10(5.8) & $6230\pm 3610$  \\
899\_340 & 8.19(8.27) & -1.7 & 23(27)  &  $110\pm 60$  \\
 883\_357 &  8.12(8.15) & -2.0 &  24(21) & $660\pm 40$  \\
 860\_359 &  8.06(8.17) & -1.8 &  4.0(5.2) & $1550\pm 150$   \\
 885\_354 &  8.21(8.26) & -1.8 & 31(34) &  $720\pm 40$ \\
 863\_348 &  7.46(7.57)& -2.0 & 35(35)  & $1620\pm 110$  \\ 
 876\_330 &  7.80(7.83) & -2.2 & 4.8(5.9)  & $740\pm 190$ \\ 
869\_328 & 6.71(6.87) & -2.8 & 31(27) &\ldots  \\
 854\_344 & 8.91(9.01)  & -1.8 & 35(35) &\ldots  \\
 854\_362 & 9.15(9.09) &  \ldots & 34(47) &\ldots \\
 846\_340 & 8.97(8.86)  & -2.9 & 1.9(3.0) &\ldots \\
\hline
\end{tabular}
\caption{Properties of spectroscopic sample of lensed galaxies presented in this paper.     The stellar masses and star formation rates are discussed in \S4.2, and 
the UV continuum slopes ($\beta$) are presented in \S4.4.   We report the best fitting stellar masses and specific star formation rates 
for models with and without nebular emission.     The latter are given in parenthesis.    
  Having only two photometric datapoints (F814W and K$_s$), the SED of 854\_362 does not permit 
determination of the UV slope, so the SFR determination assumes no dust correction.    }
\end{table}

\subsection{Low stellar masses and large specific star formation rates}

The stellar masses of our sample are listed in Table 5.   For the 
16 CIII] emitting galaxies, the  SED fitting procedure (using models with nebular emission) suggests stellar masses 
between 2.0$\times$10$^6$ M$_\odot$ and 1.4$\times$10$^{9}$ M$_\odot$.   The median stellar mass is 6.3$\times$10$^7$ 
M$_\odot$\footnote{If we instead use models with only stellar continuum,
the stellar masses are somewhat larger for select systems (see Table 5), but  the median mass  
(8.1$\times$10$^7$ M$_\odot$) of the CIII] emitting population remains very low.}, 
roughly 65-120$\times$ lower than that of L$^\star_{\rm{UV}}$ galaxies at $z\simeq 2-3$ (Shapley et al. 2005, Reddy \& Steidel 2009, 
Reddy et al. 2012) and more than a factor of ten less massive (in stars) than BX 418, the metal poor CIII] emitter 
studied in Erb et al. (2010).   

Our spectra indicate that (similar to BX 418) the physical conditions in many star forming galaxies with stellar masses 
between 10$^6$ and 10$^8$ M$_\odot$ are particularly conducive to the production of large equivalent width collisionally-excited 
lines.    Based on the  presence of CIII] emission among more massive ($\simeq 10^{9}$ M$_\odot$) 
galaxies in our sample (854\_362) and elsewhere in the literature  (Shapley et al. 2003, Erb et al. 2010, Christensen et al. 2012,
Stark et al. 2013b), we suspect that low stellar mass is not the dominant factor regulating the strength of collisionally excited emission 
lines.   More likely the connection between low stellar mass and CIII] emission is indicative of 
the dependence of yet more important properties (i.e., metallicity) on stellar mass.  

Examination of existing samples in the literature suggests that the specific star formation rate might also be an important indicator 
of large equivalent width CIII] emission.   For example, 
while the masses of M2031 (1.4$\times$10$^{9}$ M$_\odot$; Christensen et al. 2012a) and BX 418 (0.9$\times$10$^{9}$ 
M$_\odot$ for assumed Chabrier IMF; Erb et al. 2010) are not nearly 
as low as many of the galaxies in our sample, their CIII] 
equivalent widths are comparable.   In addition to having low metallicities, both galaxies stand out as having unusually 
large specific star formation rates (17 Gyr$^{-1}$ for BX 418 and 13 Gyr$^{-1}$ for M2031) among galaxies with 
moderate ($\sim 10^{9}$ M$_\odot$) stellar masses.

Similar to BX 418 and M2031, the  specific star formation rates of the extreme CIII] emitters in our sample are very large.   
The median value determined from the continuum SEDs, 27 Gyr$^{-1}$, is more than 10$\times$  that of typical 
$z\simeq 2-3$ UV-selected galaxies (e.g., Reddy et al. 2012), indicating the dwarf galaxies with powerful line emission have 
undergone a phase of unusually rapid stellar mass growth over the past 100 Myr (the timescale probed by the 
UV continuum luminosity).\footnote{This average is based on modelling with a 
minimum age of 30 Myr.   As discussed in \S4.1, lowering the minimum age will tend to increase the inferred star formation rate.
If we were to adopt a 5 Myr as the lower age threshold, we would find a median sSFR of 32 Gyr$^{-1}$.  As noted in \S4.1, the stellar 
masses could be underestimated by a factor of 2-3 if an underlying old stellar component is hidden 
by the current episode of star formation powering the extreme line emission.   But even in this case, the specific star formation 
rates are still well above average (e.g., Reddy et al. 2012).}   Galaxies caught in the midst of such an upturn in their star formation might be 
expected to exhibit prominent nebular emission lines for a variety of reasons.  First, the reservoir of ionising radiation per unit mass 
will be greater during such an active period of star formation, causing an enhanced ionisation parameter.    Second 
the ionising continuum is likely to be somewhat harder at young ages, translating into a larger electron temperature and stronger 
collisionally excited lines.    In the future, it should be possible to confirm the connection between sSFR and UV lines through 
a comparison of the H$\alpha$ equivalent width (a proxy for the sSFR over the last 10 Myr) and the CIII] equivalent width.
 
But the inference of a very large sSFR from a broadband SED does not guarantee 
extreme equivalent width UV emission lines.   For example, the sSFR of the only galaxy in 
our sample without a CIII] detection (MACS 0451-6.2) is between 23 and 35 Gyr$^{-1}$, 
comparable to many of the UV line emitting galaxies in our sample.   
 Clearly factors other than the sSFR 
(i.e., metallicity, dust content) play an important role in regulating the equivalent 
width of the UV emission lines.

An additional complication in using the derived sSFR as an indicator of whether 
or not a galaxy is likely to be an extreme UV line emitter is that the UV continuum (which we used 
to derive the SFR) only reflects activity on 100 Myr timescales.   If star formation 
has fallen off significantly in the past $\sim10$ Myr (as might be expected if star formation is bursty), then the sSFR 
derived from the stellar continuum might still be large, but the hot stars required to power 
nebular emission lines would be mostly absent.   As we will argue in \S4.4, this is 
a possible explanation for the weak nebular line emission in 869\_328, a very low 
stellar mass galaxy with a continuum-derived specific star formation rate between 
27 and 31 Gyr$^{-1}$.   

\subsection{Extreme [OIII]+H$\beta$ Equivalent Widths}

\begin{figure*}
\begin{center}
\includegraphics[width=0.98\textwidth]{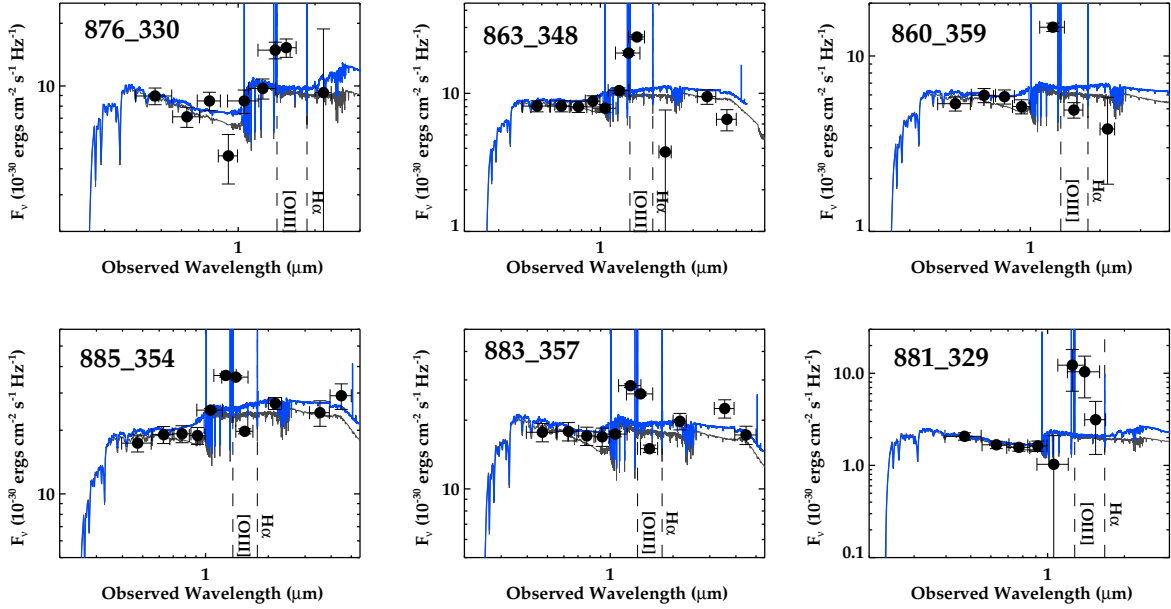}
\caption{Evidence for large EW rest-optical [OIII]+H$\beta$ emission in CIII] emitters.  The impact of [OIII]+H$\beta$ emission is clearly seen 
on the near-infrared broadband photometry.   The dashed lines denote the observed wavelengths of [OIII]$\lambda$5007 
and H$\alpha$.   We show the best-fitting population synthesis 
models with stellar continuum and nebular emission in blue.   The best fitting models with only stellar continuum  (grey) provide a much poorer fit 
to the observations.  The  The [OIII]+H$\beta$ 
equivalent widths required to explain the observed flux excesses are typically in excess of 660~\AA\ rest-frame (see Table 4).  }
\label{fig:o3hb}
\end{center}
\end{figure*}

Recent studies have identified a population of galaxies at high redshift with very large rest-optical equivalent width ($\gsim 500-1000$~\AA) emission lines.  There is 
now emerging evidence that such extreme (rest-optical) emission line galaxies (EELGs) become substantially more common at higher redshifts (e.g., Shim et al. 2012, Stark et al. 2013a, Smit et al. 2013).   The latest results at $z\gsim 6$ (Smit et al. 2013, Labb\'{e} et al. 2013) indicate that most early star forming galaxies have extremely large equivalent width rest-optical emission lines.  

Since the physics of collisionally-excited emission lines is essentially the same in the ultraviolet and optical, it seems likely 
that the physical conditions (i.e., metallicity, sSFR) which cause the extreme optical line emission  in EELGs are 
identical to those responsible for the prominent UV emission lines in our sample of lensed galaxies.   
If this is the case, we should expect the galaxies in our sample to exhibit [OIII]$\lambda$5007 emission lines with equivalent widths similar to the population of EELGs.   Given  the ubiquitous presence of large equivalent width rest-optical line emission in galaxies at $z\gsim 6$, empirical evidence 
demonstrating a firm connection between the EELG population and our extreme rest-UV line emitting galaxies would 
help build the case that rest-UV lines such as CIII] will be present in the spectra of $z\gsim 6$ galaxies.  

While efforts to measure the rest-optical emission lines of our intermediate redshift sample with near-infrared spectroscopy are now underway (\S4.5), 
many of the extreme CIII] emitting galaxies are at redshifts which place their rest-optical emission 
lines in regions of low atmospheric transmission and are thus difficult to characterise from the ground.  However if the galaxies in our sample have extremely large rest-optical equivalent widths, we should see prominent flux excesses in the HST WFC3/IR 
broadband filters which contain the emission lines.   In our previous work 
(Stark et al. 2013a, Schenker et al. 2013b) and that of other groups (e.g., van der Wel et al. 2012, Smit et al. 2013), 
it has become customary to use the observed flux excesses in specific redshift intervals to characterise 
rest-optical line emission line equivalent widths.   

We will focus our rest-optical line analysis on galaxies in the Abell 1689 field.   With 
deep WFC3/IR imaging in four filters (Y$_{105}$, J$_{125}$, J$_{140}$, H$_{160}$), we can 
identify large equivalent width [OIII] emission via its influence on the  flux density in either the J$_{125}$-band 
(for galaxies at $1.18<z<1.80$), the J$_{140}$-band ($1.38<z<2.20$), or the H$_{160}$-band ($1.78<z<2.37$).   
Throughout much of these redshift intervals, the observed flux excesses will also be affected by H$\beta$ 
emission, and hence the inferred equivalent widths will reflect the [OIII]+H$\beta$ line strengths.  
Seven of the CIII] emitters toward Abell 1689 fall in one of the redshift intervals noted above.    It is immediately clear from 
examination of the SEDs in Figure 5 that the observed flux in the 
bandpasses which are contaminated by [OIII]+H$\beta$ are substantially in excess of that expected from stellar 
continuum emission.   Following 
the techniques we used in Stark et al. (2013a), we compute the line flux and equivalent width required to produce  
the flux excess.   

The median (rest-frame) [OIII]+H$\beta$ equivalent widths required to explain the flux excess is 740~\AA\ (Table 4).   Six of  
seven galaxies in the appropriate redshift window have inferred rest-frame [OIII]+H$\beta$ equivalent widths in excess of 660~\AA, similar to many of the 
rest-optically selected EELGs at similar redshifts (e.g., Atek et al. 2011, van der Wel et al. 2011, Maseda et al. 2013).   The 
physical link between the UV and optical line emitters is not surprising, as we outlined above.   If the ubiquity of large 
equivalent width optical emission lines is confirmed with larger samples at $z\gsim 4$ (e.g., Stark et al. 2013a, Smit et al. 2013, Schenker et al. 2013, Holden et al. 2013, Labb\'{e} et al. 2013), the 
connection would bolster  confidence that ultraviolet emission lines such as CIII] or OIII] might be visible in suitably deep 
spectroscopic exposures.  

\subsection{Blue UV continuum slopes}

The UV-continuum slope (parameterized as $\beta$ where $f_\lambda \propto \lambda^\beta$) 
provides a unique constraint on the dust extinction in high-redshift UV-selected galaxies (e.g., Reddy et al. 
2012a) and is additionally affected by the metallicity and age of the stellar population.   Over the last several years, 
considerable effort has been focused on determining the distribution of UV-continuum slopes in galaxies at very high redshifts 
(e.g., Wilkins et al. 2011, Finkelstein et al. 2012, Rogers et al. 2013, 
Dunlop et al. 2013, Bouwens et al. 2012, 2013).     
These studies reveal that $z\simeq 7$ galaxies have  blue UV slopes ($\beta \simeq  -2.0$ to  $\beta \simeq  -2.5$), consistent with 
minimal dust reddening.   Comparison to lower redshift samples 
reveals mild redshift evolution, with galaxies at $z\simeq 7$ bluer than those of similar UV luminosity 
at $z\simeq 2.5$ by $\Delta\beta\simeq 0.3$  (Bouwens et al. 2013).   
While some controversy remains at the highest redshifts, the UV slopes show a trend 
toward bluer colours at lower UV luminosities in galaxies at $z\simeq 4-5$ (e.g., Wilkins et al. 2011, 
Bouwens et al. 2013), as would be expected if low luminosities galaxies are metal poor with little reddening from 
dust.

Here we measure the UV slopes of our lensed galaxy sample and investigate whether there is a connection 
with the equivalent width of CIII] emission.   UV slopes are derived using the flux information from filters spanning 
rest-frame wavelengths of 1300~\AA\ and 2500~\AA.\     Given the spread of redshifts within our sample (and the 
different filter deployment in the three cluster fields), the exact 
filters used vary from galaxy to galaxy.   We take care to exclude filters which are affected by the 
Ly$\alpha$ forest or Ly$\alpha$ emission.   The metal UV lines do not have large enough equivalent width to 
significantly affect the broadband flux measurements.   UV slopes are derived for each source in our spectroscopic 
sample using a simple least squares minimisation procedure.    

The  UV slopes are provided in Table 5.     The 
median value ($\beta \simeq -2.2$)  is considerably bluer than that of  more 
luminous systems at similar redshifts (e.g., Reddy et al. 2012a), 
likely reflecting the reduced dust content, 
lower metallicities and younger ages of the low luminosity galaxies in our sample. 
Comparison of our 
sample with the more luminous systems in Shapley et al. (2003) reveals that galaxies with blue colours ($\beta \lsim -2$) do tend to have larger  
equivalent width CIII] emission than galaxies with redder UV slopes  ($\beta \gsim -1.3$).   The importance of UV slope as an 
indicator of CIII] equivalent width is made evident by MACS 0451-6.2, the only galaxy in our sample without CIII] emission.   
While  its  specific star formation rate is as large as many of the most prominent line emitters that we consider in this paper, MACS 0451-6.2 
stands out as the reddest galaxy ($\beta = -1.2$) in our sample.   As is evident from Figure 6, the red UV colour  places 
MACS 0451-6.2 in a regime where CIII] equivalent widths are generally very  low.  

The dependence of CIII] equivalent width on UV colour  is  
to be expected.   Galaxies with UV slopes bluer than $\beta \sim -2$ (indicative of little to no
reddening from dust) are likely to be reasonably metal poor.  Given the larger electron temperatures 
associated with ionised gas in galaxies with moderately low metallicities (see \S4.5 for more discussion), 
star forming systems with very blue UV colours are prime candidates for having large equivalent width emission lines in the 
ultraviolet and optical.    The low dust covering fraction of galaxies with very blue UV colours 
may also play an important role.   If HII regions are more strongly attenuated than 
the stellar continuum emission (see discussion in \S4.1), then the equivalent widths of UV 
emission lines will be weakened in galaxies with reddened continuum.   

Given that the UV continuum slope is one of the few quantities that can be  measured at 
$z\gsim 7$, the relationship between colour and CIII] equivalent width  will be  
crucial for assessing the feasibility of pursuing CIII] in reionization-era galaxies and will provide a useful 
baseline for interpreting the results.    With most of the star forming population at $z\gsim 7$ exhibiting  blue UV slopes, it seems likely that 
large equivalent width CIII] emission should be much more common in early star forming galaxies.  

But as is clear from the  scatter  in the UV slopes in Table 5,  galaxies with blue UV colours 
do not always have extreme UV line emission.   This point is nicely illustrated 
by the spectrum of 869\_328, one of the lowest stellar mass (5.1$\times$10$^6$ M$_\odot$) galaxies 
in our sample.   While the UV slope of 869\_328 is among the bluest in our sample ($\beta \simeq -2.8$),  
the equivalent width of CIII]  (1.8~\AA) and Ly$\alpha$ (4.3~\AA) are both found to be very low (although 
there is extended Ly$\alpha$ emission spanning $\simeq 25$\arcsec, Richard et al. 2014 in preparation)  As is 
clear from the SED (Figure 6), this galaxy appears very young, with an sSFR of between 27 and 31 Gyr$^{-1}$.   
Galaxies with such large sSFR are commonly accompanied by very prominent nebular emission.  
But the nebular + stellar continuum model struggles to reproduce the observed SED.   Not only are the 
emission lines much weaker than expected, but the impact of nebular continuum causes the 
UV slope of the best fitting model (blue curve overlaid in Figure 6) to be considerably redder than observed.   

The physical origin of the weak nebular line and nebular continuum emission in 869\_328 is not clear.  
One possibility is that the bulk of the ionising radiation is escaping the galaxy.   It is also possible that the  massive O and early B stars 
required to power  nebular emission are mostly absent.   Such a situation could occur from bursty star formation or 
stochastic sampling of the IMF.    In the local volume, the lowest luminosity star forming galaxies often exhibit very  
low ratios of H$\alpha$ to far UV continuum (e.g., Lee et al. 2009).   It is conceivable that 869\_328, one of the lowest 
mass galaxies in our sample, is part of an analogous population at high redshift.   But clearly some caution 
must be exercised in interpreting this target given the limited nature of the data in hand.  Rest-optical emission lines (in particular H$\alpha$) will soon provide a considerably more reliable picture 
of the nature of this source.

\begin{figure}
\begin{center}
\includegraphics[width=0.47\textwidth]{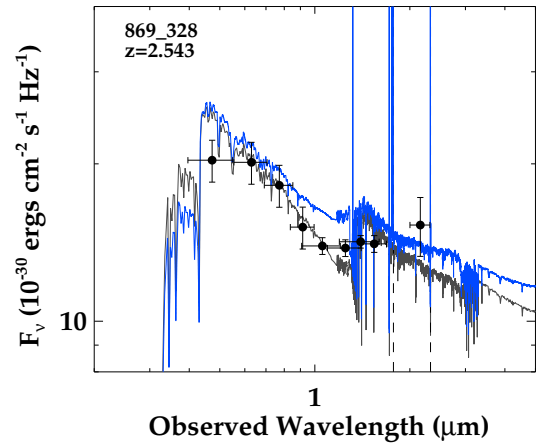}
\caption{Example of low stellar mass (5.1$\times$10$^{6}$ M$_\odot$) galaxy in our sample with low 
equivalent width UV nebular emission lines (W$_{\rm{CIII]}}$=1.8~\AA).  The best fitting stellar continuum 
(dark grey) and nebular+stellar models (blue) are overlaid.   The broadband SED is among the 
bluest in our sample ($\beta = -2.8$) and the sSFR reveals significant stellar mass growth in the past 100 Myr. 
The UV slope of the best-fitting nebular+stellar model is considerably redder than observed owing to 
nebular continuum emission.  The stellar continuum models  provide a much better fit 
to the data.  } 
\label{fig:869328}
\end{center}
\end{figure}

\subsection{Low Gas-Phase Metallicity}

As we emphasized at the outset of \S4, the gas-phase metallicity plays an important role in governing the strength of 
UV emission lines such as CIII] and OIII].    The large electron temperature that arises due to inefficient 
cooling in moderately low metallicity gas tends to increase 
the equivalent width of collisionally-excited emission lines, while the harder ionizing output from low metallicity stellar populations 
is likely to boost the output of ionising photons.   Perhaps not surprisingly, the low mass star forming galaxies 
known at high redshift with large equivalent width CIII] emission (e.g., Erb et al. 2010, Christensen et al. 2012b, James et al. 2013) 
tend to have optical 
emission line ratios consistent with moderately metal poor ($\lsim  0.2$ Z$_\odot$) ionised gas (an exception is the CIII] emitter 
reported in Bayliss 
et al. 2013 with 0.5 Z$_\odot$).  
Here we characterise the gas-phase chemical abundances of  a
small subset of galaxies in our sample.  

\begin{figure}
\begin{center}
\includegraphics[width=0.49\textwidth]{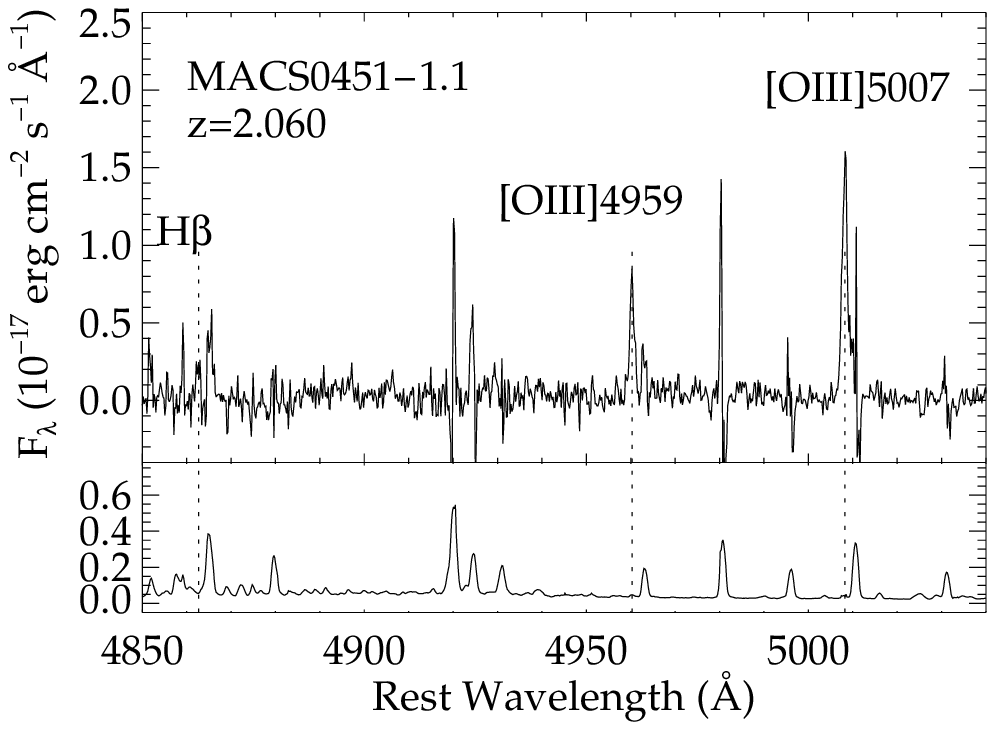}
\includegraphics[width=0.49\textwidth]{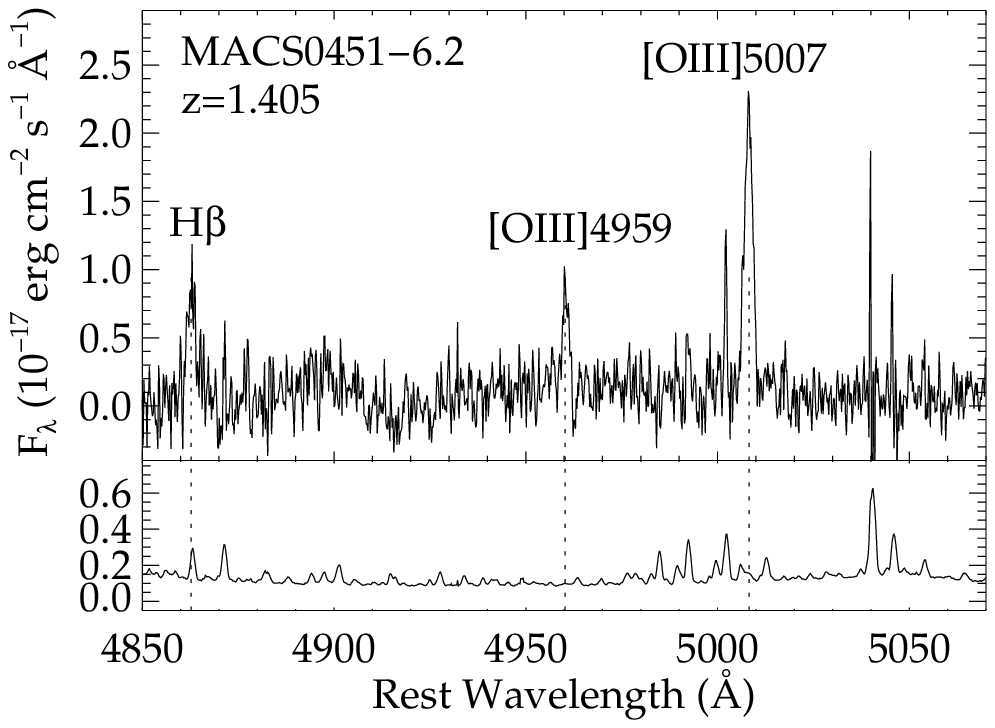}
\caption{Magellan/FIRE near-infrared spectrum of MACS 0451-1.1 (top panel) and MACS 0451-6.2 (bottom panel).  
The 1$\sigma$ noise spectrum is presented underneath the object spectrum in both panels.   Emission line fluxes 
are listed in Table 6.
 }
\label{fig:firespec}
\end{center}
\end{figure}

\subsubsection{Characterisation of optical emission lines}

We calculate nebular redshifts and line fluxes through Gaussian fits to the emission lines detected with FIRE and 
XSHOOTER.   
To  determine the uncertainty 
in the line fluxes and redshifts, we generate 1000 realisations of the spectra by perturbing the flux at 
each wavelength by an amount consistent with the error spectrum.   We fit the line fluxes and redshifts of 
each of the mock spectra.   The standard deviation of the flux and redshift distribution is then used to 
compute 1$\sigma$ uncertainties in each of these quantities.    In the remainder of this sub-section, we 
provide brief descriptions of the emission lines detected in each of the five galaxies targeted.

The XSHOOTER spectrum of 860\_359 shows strong emission from H$\beta$, [OIII]$\lambda$4959, [OIII]$\lambda$5007, 
and H$\alpha$ (see Amanullah et al. 2011).   Line fluxes are listed in Table 6.  The flux ratio of H$\alpha$ and 
H$\beta$ ($3.0 \pm 0.1$) suggests that the rest-optical 
emission lines are not strongly reddened by dust.   The redshift of the galaxy places [OIII]$\lambda$5007 at 
an observed wavelength (1.353 $\mu$m) strongly attenuated by telluric absorption.   While the line is significantly 
detected, the flux measurement is fairly uncertain owing to the low atmospheric transmission.   We thus infer the 
[OIII]$\lambda$5007 flux using our measurement of [OIII]$\lambda$4959 (which is located at a wavelength less 
strongly attenuated by telluric absorption) and the theoretical [OIII] 5007/4959 flux ratio (set by statistical weights) of 2.98.   
[NII]$\lambda$6584 is not detected with significance.   A 2$\sigma$ upper limit is listed in Table 6. 
The [OII] emission line is in a noisy region of the XSHOOTER spectrum (in between the peak sensitivity of the visible 
and near-infrared arm) resulting in coarse upper limits on the flux.

MACS 0451-1.1 was observed with FIRE on 16 Feb 2012 and 29 Oct 2012.   The 16 Feb 2012  observation 
of MACS 0451-1.1 was  oriented perpendicular 
to the extended arc to ensure optimal sky subtraction (similar to the VLT/FORS orientation), covering only the central region.   
The position angle adopted 
on 29 Oct 2012 was chosen  to include a larger fraction of the lensed galaxy.   In both spectra, we confidently detect 
H$\alpha$, [OIII]$\lambda$4959, and [OIII]$\lambda$5007 (Figure 7; see Table 6 for flux measurements).     
Sky subtraction residuals prevent useful constraints on [NII]$\lambda$6584, and [OII]  is not detected.    
We see H$\beta$ confidently in the 29 Oct 2012 spectrum.   The flux ratio of  H$\alpha$ and H$\beta$ ($2.6\pm 0.3$) suggests that 
the emission lines are not  strongly affected by reddening.   In the 16 Feb 2012 spectrum, the formal 
significance of the H$\beta$ detection is much lower (S/N=1.6).   Rather than introduce a noisy H$\beta$ 
measurement into  analysis of the 16 Feb 2012 spectrum, we follow the approach taken in similar  cases 
(e.g., Hainline 
et al. 2009) of estimating the H$\beta$ flux by adjusting the measured H$\alpha$ flux assuming the ratio for case B 
hydrogen recombination and accounting for reddening based on the E(B-V) inferred from the SED.   We assume 
an intrinsic H$\alpha$/H$\beta$ ratio of 2.81 (the average of the case B ratios expected for gas with T$_e$=
1.0$\times$10$^4$ K and 2.0$\times$10$^4$ K; Osterbrock 1989) and assume no reddening.   The estimated value is listed in Table 6.

The near-IR spectrum of MACS 0451-6.2  reveals 
 detections of many of the strongest rest-optical emission lines (Figure 7) including [OII], H$\beta$, [OIII], 
 and H$\alpha$.  The [NII]$\lambda$6584 emission line is at the edge of a sky line but is significantly 
detected.   By fitting the half of the line which is unaffected by sky residuals, we infer a total line flux 
of 3.1$\times$10$^{-17}$ erg cm$^{-2}$ s$^{-1}$.    We also detect a faint emission feature at the location 
of [NII]$\lambda$6548 with a line flux of 9.2$\times$10$^{-18}$ erg cm$^{-2}$ s$^{-1}$.   Assuming a 
[NII] 6583/6548 doublet flux ratio of 2.93 (as set by the statistical weights of the energy levels), we derive an [NII]$\lambda$6584 flux of 
2.7$\times$10$^{-17}$ erg cm$^{-2}$ s$^{-1}$.  
We use the mean value of the two  [NII]$\lambda$6584 flux estimates in Table 6.  
 The measured Balmer decrement (H$\alpha$/H$\beta \simeq 3.6$) indicates that dust reddening moderately 
affects the rest-optical emission line flux ratios (consistent with a selective extinction of 
E(B-V)$_{\rm{gas}}$=0.20 following Calzetti et al. 2000) as expected based on the red UV continuum slope ($\beta\simeq -1.2$) inferred from the SED.

The final two galaxies remaining in Table 6 are MACS 0451-3.1 and 899\_340.  The FIRE spectrum of 
MACS 0451-3.1 contains strong 
[OIII]$\lambda$5007 emission, indicating a nebular redshift of $z=1.9043$.    At this redshift, 
most of the other strong rest-optical emission lines are in regions of low atmospheric transmission 
and are not detected.    The  FIRE spectrum of 899\_340 exhibits [OIII]$\lambda$5007 and H$\alpha$, indicating a 
nebular redshift of $z=1.600$.   Both 
[OIII]$\lambda$4959 and [NII]$\lambda$6584 are obscured by sky lines.  The [OII]  and 
H$\beta$ lines are located in clean regions but are not detected.   Line fluxes (and 
2$\sigma$ upper limits) are listed in Table 6.

\begin{table*}
\begin{tabular}{lcccccc}
\hline 
& MACS 0451 & MACS 0451 & MAC S0451  & MACS 0451 & Abell 1689 & Abell 1689 \\ \hline 
& 1.1$a$ & 1.1$b$ & 6.2 & 3.1 & 899\_340 & 860\_359 \\ \hline 
 \hline 
z$_{neb}$ & 2.0596 & 2.0596 & 1.4048 & 1.9043 & $1.5996$ &  1.7024 \\
F$_{\rm{[OII]\lambda3726}}$ & $<0.6$ & $<$0.6 &$2.5\pm 0.7$ &$<0.9$ & $<3.1$ & $< 1.9$ \\
F$_{\rm{[OII]\lambda3729}}$ & \ldots & $<$0.6 &$3.2 \pm 1.0$  & $<1.1$ & $<3.7$ & $<1.7$ \\
F$_{\rm{H\beta}}$ & $1.3^\dagger\pm 0.5$ & $1.9\pm 0.2$ & $4.9 \pm 0.4$ & \ldots & $<2.5$ & $3.1\pm0.1$  \\
F$_{\rm{[OIII]\lambda4959}}$ & $2.9\pm 0.1$  & $2.6\pm0.1$ & $5.1\pm 0.3 $ & \ldots  &\ldots  & $6.3\pm 0.1$\\
F$_{\rm{[OIII]\lambda5007}}$&  $7.8\pm 0.2$& $7.5\pm0.1$ & $13.7\pm 0.5$  & $3.1\pm 0.4$ &$9.0 \pm 1.2$ & $18.6^\dagger\pm 0.1$ \\
F$_{\rm{H\alpha}}$  & $3.6\pm0.2$ & $4.9 \pm 0.3$ & $17.7\pm 0.3$& \ldots   & $7.0 \pm 0.9$ & $9.2\pm 0.1$ \\
F$_{\rm{[NII]\lambda6584}}$ & \ldots & \ldots  &$2.9^\dagger\pm 0.3$ & \ldots & \ldots  & $\lsim 0.2$ \\
\hline
\end{tabular}
\caption{Rest-optical emission line flux in units of 10$^{-17}$ erg cm$^{-2}$ s$^{-1}$.   In cases where no 
emission is detected, we list 2$\sigma$ upper limits.   The fluxes listed above do not include  the fraction of the lensed galaxy 
which is not sampled by the slit.   MACS 0451-1.1 was observed using two separate position angles with 
Magellan/FIRE.   The epoch 1 observations (16 Feb 2012) are denoted 1.1a while the epoch 2 observations 
(29 Oct 2012) are listed as 1.1b.  Emission lines flagged with a dagger symbol ($\dagger$)  are indirect measurements owing 
to sky subtraction residuals, telluric absorption or low S/N; for details see \S4.5.1.}
\end{table*}

\subsubsection{Metallicity estimates}

The rest-optical emission lines discussed in \S4.5.1 allow us to constrain the 
gas-phase metallicity of three galaxies (MACS 0451-1.1, MACS 0451-6.2, 860\_359) in our sample.    Since the spectra are not deep enough to 
detect the electron temperature (T$_e$) sensitive [OIII]$\lambda$4363 emission line, we estimate 
oxygen abundances using standard strong-line abundance indicators 
based on the relative strength of strong collisionally-excited emission lines ([OII], [OIII], [NII]) and 
hydrogen recombination lines (H$\alpha$, H$\beta$) in the rest-frame optical.
In the following, we will consider the R$_{\rm{23}}$ index  
(R$_{\rm{23}}\equiv$ ([OII] + [OIII]) / H$\beta$),  the N2 index ($\rm{N2\equiv  log \{[NII]\lambda}$6584  / 
H$\alpha$)\}, and the O3N2 index (O3N$2\equiv \rm{log \{[OII]}$/H$\beta$)/ ([NII]/H$\alpha$)\}.   While 
each of these indices can provide an estimate of the oxygen abundance,  there are  
systematic offsets in the absolute abundance scales of the different calibrations.   We account for these 
using the conversions presented in Kewley \& Ellison (2008).   To reliably map these strong line indicators to an oxygen
abundance, care must be taken to assess the ionisation state of the gas.    This is usually done 
through constraining the ionisation parameter, U, defined as the ratio of the 
surface flux of hydrogen ionising photons and density of hydrogen atoms.  For this purpose, we will also   
calculate the ratio of the [OIII] and [OII] lines (O$_{32} \equiv $ \{[OIII]$\lambda$4959+[OIII]$\lambda$5007\} / 
[OII]$\lambda$3727).

We will make use of the Pettini \& Pagel (2004) calibration (PP04) of the N2 and 
O3N2 index.   For the R$_{\rm{23}}$ index, we follow the  
iterative approach discussed in Kobulnicky  \& Kewley (2004) (KK04), which is based on calibrations derived 
from the photoionisation models of Kewley \& Dopita (2002).      The R$_{\rm{23}}$ index is well known to be double valued 
with metallicity, requiring external constraints to establish whether the galaxy is 
low metallicity (and thus on the ``lower branch") or high metallicity (on the ``upper branch").    For each galaxy we consider below, 
we will discuss whether the 
upper branch or lower branch is appropriate.   
As discussed above, the relationship between the 
measured R$_{\rm{23}}$ value   and the oxygen abundance, O/H, is dependent on the ionisation parameter.   While the 
O$_{32}$ diagnostic constrains the ionisation parameter, the relationship between the two quantities depends 
on the metallicity.   We therefore must adopt an iterative scheme to ensure convergence in U and O/H.    
Using equation 13 of KK04, we calculate U from our measurement 
of O$_{32}$ and an initial guess at O/H.    We then derive the oxygen abundance by inputting 
the inferred ionisation parameter and measured R$_{\rm{23}}$ index into the lower or upper branch R$_{\rm{23}}$ calibration 
of KK04.   If the final inferred metallicity is significantly different from our 
initial guess, we re-do the calculation inserting the derived value of O/H into the ionisation parameter calculation. 
Several cycles are typically required before convergence is reached.   

The first system we consider is 860\_359, one of the the extreme CIII] emitting  galaxies in our sample. 
Based on the flux ratios in Table 6, we measure $\rm{N2<-1.7}$ and $\rm{O3N2>2.4}$.  Following the 
calibrations of PP04, these measurements suggest an oxygen abundance of 12 + log O/H $<  7.9$ for 
N2 and O3N2.   Assuming the solar abundance value 
of 12+log O/H = 8.66 (Asplund et al.  2009), these observations suggest that the typical metallicity is less 
than 0.2 Z$_\odot$.   The ratio of [NII] and H$\alpha$ places the galaxy on the lower branch of the 
R$_{\rm{23}}$ - O/H relationship (see e.g., KK04).   Taking into consideration the 2$\sigma$ limit 
on the [OII] flux, we find $8.0 < \rm{R_{23}} < 9.2$ and O$_{\rm{32}} > 6.9$.   Applying the iterative KK04 procedure 
described above, we find that these constraints translate into an upper limit on the 
oxygen abundance of 12+log O/H $ < $8.3.   Applying the mapping derived in Kewley \& Ellison (2008), 
we find that the KK04 abundance limit corresponds to 12+log O/H$ < $8.2 on the PP04 N2 abundance scale.   
Both calibrations suggest that the ionised gas in 860\_359 is moderately metal poor with an oxygen abundance 
no greater than 0.2 - 0.3 Z$_\odot$.

The second system we examine is MACS 0451-1.1, a moderate equivalent width CIII] emitter with 
low stellar mass (4.9-6.3$\times$10$^{7}$ M$_\odot$), large specific star formation rate (28-35 Gyr$^{-1}$) 
and blue UV slope ($\beta=-2.7$).    
As explained in \S 4.5.1, the galaxy was observed with two different position angles 
with FIRE.  
The non-detection of [OII] in both epochs bounds the R$_{\rm{23}}$ index to the range 8.4-9.3
(16 Feb 2012) and 5.3-5.9 (29 Oct 2012).     We find that O$_{\rm{32}} > 8.9$ (16 Feb 2012)  and 
O$_{\rm{32}} > 8.4$ (29 Oct 2012).   Based on the Balmer decrement measurement (see discussion 
in \S4.5.1), we do not apply reddening corrections to the line ratios.   Without a robust measurement 
on [NII], we cannot reliably place MACS 0451-1.1 on the upper or lower branch.   However most available 
evidence for MACS 0451-1.1 (low luminosity, low 
stellar mass, blue UV continuum slope, detection of T$_e$-sensitive line OIII]$\lambda$1666 emission line) suggests that the galaxy  resides on the 
lower branch.    Following the iterative KK04 approach (and considering a range of O$_{\rm{32}}$ and R$_{\rm{23}}$ values consistent 
with our [OII] flux limit), we find 12+log O/H $ \lsim 8.0$ (29 Oct 2012) and 12+log O/H $ \lsim 8.3$ for 
(16 Feb 2012).
 In the more unlikely case that MACS 0451-1.1 
lies on the upper branch, the KK04 parameterisation implies  12+log O/H $\gsim 8.8$ (29 Oct 2012) and 
12+log O/H $\gsim 8.5$ (16 Feb 2012).\footnote{ Using the relations presented in Kewley \& Ellison (2008), we convert the 
KK04 oxygen abundances to the absolute abundance scale of the PP04 N2 calibration.    The 
lower branch measurements are similar for both calibrations.   
But we find that the upper branch abundance metallicities are 
slightly lower on the PP04 N2 scale: 12+log O/H$ \gsim 8.4$ (29 Oct 2012 spectrum) and 12+log O/H$\gsim $8.2 (16 Feb 2012 spectrum).}   
The differences in the metallicities between the two epochs may 
reflect physical differences in the oxygen abundance probed by the two slit positions.  Integral field observations 
are required to reliably map the metallicity gradient and investigate whether shocks or AGN might drive  line ratio 
variations across the galaxy (e.g., Jones et al. 2012, Yuan et al. 2013).    But regardless of the variation in metallicity across the source, 
the integrated measurements derived here suggest that the ionised gas in MACS 0451-1.1 is likely to be 
moderately metal poor ($\lsim 0.2-0.4$ Z$_\odot$).

The third and final galaxy we examine is MACS 0451-6.2.   As we have already discussed, MACS 0451-6.2 
has no detectable CIII] emission and is 
a reasonably massive (1.2$\times$10$^{9}$ M$_\odot$ in stars) and red  ($\beta=-1.2$) galaxy for our sample.  The relative fluxes of [NII], H$\alpha$, and [OIII]$\lambda$5007 indicate that N2 = $-0.8$ and 
O3N2=1.7.  Using the PP04 calibration, we derive oxygen abundances of 12 + log (O/H) = 8.4 and 8.3 from the N2 and O3N2 calibrations, respectively.   Again applying the Kewley \& Ellison (2008) conversions, we find that these  abundances correspond to 12+log (O/H) = 8.8 and 8.7 on the KK04 abundance scale.   The measured N2 value 
places this galaxy  on the upper branch of 
the R$_{\rm{23}}$ - O/H relationship.    Without any reddening correction, we find that R$_{\rm{23}}$ = 5.0.   
 After correcting the line fluxes for reddening, we derive a slightly larger value of 
R$_{\rm{23}}$ = 5.2.      Following the KK04 iterative procedure described above (although now 
using the upper branch calibration), we infer an oxygen abundance of 12 + log (O/H) = 8.8 from the 
de-reddened flux ratios, consistent with expectations from the N2 and O3N2 indices.     
 
The metallicity estimates calculated above are consistent with the general picture proposed at the outset of 
\S4.   The  two galaxies considered above with large equivalent width CIII] emission appear to have moderately metal poor 
($\lsim 0.2-$0.4~Z$_\odot$) gas.   At these metallicities, the electron temperature of ionised gas is 
elevated, increasing the  strength of collisionally excited lines.    Prominent line emission is likely less common in 
more metal rich galaxies.    Indeed, the metallicity of the only galaxy in our 
sample without CIII] emission (MACS 0451-6.2) is noticeably larger (0.4-0.5 Z$_\odot$ on the PP04 scale) 
than that of the low luminosity galaxies with CIII] detections.  
 
\subsection{Gas-phase C/O ratio}

For CIII] to be a viable redshift indicator at $z\gsim 6$, carbon must be deposited into the ISM on reasonably short 
timescales.     While the presence of large equivalent width [OIII]+H$\beta$ emission in $z\simeq 7-8$ galaxies 
(e.g., Labb\'{e} et al. 2013, Smit et al. 2013) requires a non-negligible gas-phase abundance of oxygen at early times,  it could take 
much longer for carbon to reach the abundance required to produce prominent CIII] emission.    Before we 
can reliably motivate the use of CIII] as a probe of $z\gsim 6$ galaxies, we must therefore understand the physics regulating the relative 
abundance of carbon and oxygen in the ionised gas of early star forming systems.

Studies of the carbon-to-oxygen (C/O) ratio in galactic stars (Bensby \& Feltzing 2006, Fabbian et al. 2009), nearby HII regions 
(e.g., Garnett et al. 1995, 1997; Kobulnicky \& Skillman 1998, Garnett et al. 1999 ), damped Ly$\alpha$ emitters (Cooke et al. 2011), 
and high redshift galaxies (Shapley et al. 2003, Erb et al. 2010) demonstrate that the C/O ratio increases with increasing 
O/H for 12+log(O/H) $>$ 7.7.   
As discussed at length in earlier studies, the physical origin of this trend is likely associated with the  metallicity dependence of winds from massive rotating stars, along with the delayed release of carbon from lower mass stars (e.g., 
Henry et al. 2000, Akerman et al. 2004).   

Given the low metallicity  of dwarf galaxies in our sample (\S4.5), the framework described above suggests 
that carbon is likely to be substantially under-abundant with respect to oxygen.    We can quantify the C/O ratio  by constraining 
the flux ratio of OIII]$\lambda\lambda$1661,1666 to CIII] $\lambda \lambda$1907,1909 (e.g., Garnett et al. 1995, 
Shapley et al. 2003, Erb et al. 2010).    We will use the recent parameterisation 
presented in Erb et al. (2010):
\begin{equation}
\rm{\frac{C^{+2}}{O^{+2}} = 0.15 e^{-1.054/t} \frac{I(CIII] \lambda \lambda1907,1909)}{I(OIII]\lambda\lambda1661,1666)}},
\end{equation} 
where t=T$_e$/10$^4$ K.    The next step is to convert the measured C$^{+2}$/O$^{+2}$ ratio to a C/O ratio. 
Since O$^{+2}$ has a higher ionisation potential than C$^{+2}$ (54.9 eV versus 47.9 eV), carbon will be triply ionised while oxygen 
is still primarily in the O$^{+2}$ state.   Hence an observed trend toward lower C$^{+2}$/O$^{+2}$ ratios may result from 
either an increase in the ionising flux or a reduction in the C/O ratio.   To break this degeneracy, we must account for the ionisation 
state of the gas.   This is done by multiplying the observationally-inferred C$^{+2}$/O$^{+2}$ ratio by an ionisation correction 
factor (ICF) defined as the ratio of the volume fractions  oxygen in O$^{+2}$ and carbon in C$^{+2}$:    

\begin{equation}
\rm{ \frac{C}{O} =  \frac{C^{+2}}{O^{+2}}} \times ICF,
\end{equation}
The ICF can be inferred through constraints on the ionising spectrum through ionisation parameter diagnostics 
such as O$_{32}$.   Using a suite of CLOUDY v08.00 photoionisation models (Ferland et al. 1998),  Erb et al. 
(2010) demonstrated that for stellar populations with ionisation parameters in the range $\rm{-2.5 \lsim~log 
U~\lsim -2.0}$, the ICF is close to unity.   As the ionisation parameter increases to $\rm{log~U\simeq -1.0}$, the ICF 
is predicted to be between 1.4 and 1.8.  

We will focus our analysis on the dwarf star forming galaxies with the highest quality spectra 
(i.e. those listed in Table 3), which  limits our analysis to the most extreme equivalents width CIII] emitters.   
Two of the galaxies (863\_348, MACS 0451-1.1) in Table 4 have detections of  CIII] and both components of the OIII]$\lambda\lambda$1661,
1666 doublet.   Another two galaxies (876\_330, 860\_359) have detections of OIII]$\lambda$1666 and CIII]$\lambda$1908 
but no detection of  OIII]$\lambda$1661.  
Assuming an electron temperature of T$_e$=15,000 K (consistent with similarly metal poor galaxies studied in 
Erb et al. 2010 and Christensen et al. 2012b), we find that log~C$^{+2}$/O$^{+2}$ = $-1.0$ for 863\_348 and  log~C$^{+2}$/O$^{+2}$ =$-0.8$ for MACS 
0451-1.1.    For 876\_330 and 860\_359, the observed line ratios indicate that  $-1.1< ~$log C$^{+2}$/O$^{+2}< -0.8$ and $\rm{-0.8<~ log~C^{+2}/O^{+2} <-0.6}$, 
respectively.    The presence of CIV emission in 863\_348 and 860\_359 indicates that a significant component of 
carbon is triply ionised and that the ICF will certainly be greater than unity.    Allowing the ICF to be between 1.2 and 2.0, we 
find $\rm{-1.0 < log~C/O < -0.7}$ for 863\_348 and $\rm{-0.8 < log~C/O < -0.5}$ for MACS 0451-1.1.    Applying the same assumptions 
to 876\_330 and 860\_359, we find  $\rm{-1.0 < log~C/O < -0.5}$ and  $\rm{-0.8 < log~C/O < -0.3}$.

The C/O ratios of the metal poor dwarf star forming galaxies in our sample are much lower than the solar C/O ratio ($\rm{log~C/O = -0.26}$) and the C/O ratio in Orion ($\rm{log~C/O = -0.21}$; Esteban et al. 2004) but are similar to those of other metal poor high redshift galaxies (e.g., Erb et al. 2010, Christensen et al. 2012b, 
James et al. 2013), consistent with the framework described above in 
which the trend of C/O with O/H (for galaxies with $\rm{[O/H] > -1}$) is set by metallicity-dependent winds in massive stars.  
Based on the ultraviolet spectra of our sample, it is clear that galaxies with low C/O ratios can still have very large equivalent width 
CIII] and CIV emission.   While the reduced carbon abundance certainly impacts the observed CIII] line strengths, the equivalent 
widths remain very large owing to other factors (metal poor gas, young and low metallicity stars, large ionisation parameter) which 
typically accompany galaxies with low C/O ratios.  

\begin{table*}
\begin{tabular}{lrrrr}
\hline
Property & 876\_330 & 863\_348 & 860\_359 & MACS 0451-1.1 \\
 \hline
$\log{U}$   & $-2.16_{- 0.32}^{+ 0.27}$ &$ -1.84_{- 0.21}^{+ 0.15}$ & $-2.13_{- 0.16}^{+ 0.16}$ &          $-1.97_{- 0.30}^{+ 0.28}$        \\
$12+\log{(\mathrm{O/H})}$   & $ 7.74_{- 0.56}^{+ 0.27}$ &$  7.82_{- 0.53}^{+ 0.10}$ & $ 7.79_{- 0.46}^{+ 0.19}$ &          $ 7.29_{- 0.22}^{+ 0.58}$        \\
$\log(\mathrm{C/O})$   & $-0.68_{- 0.17}^{+ 0.14}$ &$ -0.74_{- 0.08}^{+ 0.08}$ & $-0.58_{- 0.08}^{+ 0.07}$ &          $-0.71_{- 0.09}^{+ 0.13}$        \\
$\log(\mathrm{age/yr})$   & $ 7.55_{- 0.48}^{+ 0.79}$ &$  6.75_{- 0.52}^{+ 0.52}$ & $ 7.30_{- 0.34}^{+ 0.56}$ &          $ 7.71_{- 0.57}^{+ 0.74}$        \\
$\log \{W(\mathrm{[OIII]4959,5007+H\beta})\}$   & $ 2.84_{- 0.28}^{+ 0.25}$ &$  3.18_{- 0.18}^{+ 0.09}$ & $ 3.00_{- 0.19}^{+ 0.18}$ &          $ 2.78_{- 0.23}^{+ 0.24}$        \\
$\mu$   & $ 0.70_{- 0.39}^{+ 0.18}$ &$  0.78_{- 0.35}^{+ 0.12}$ & $ 0.77_{- 0.38}^{+ 0.13}$ &          $ 0.70_{- 0.39}^{+ 0.18}$        \\
$\hat{\tau}_V$   & $ 0.20_{- 0.20}^{+ 0.40}$ &$  0.01_{- 0.18}^{+ 0.22}$ & $ 0.14_{- 0.19}^{+ 0.42}$ &          $ 0.19_{- 0.18}^{+ 0.38}$        \\
HeII/CIII]     &       0.006      &        0.005     &          0.004         &     0.002   \\
CIV/CIII]       &    0.438        &      0.892      &          0.404    &         1.113  \\
EW(CIII])      &    9.597   &         10.556        &      12.302     &        6.715  \\
     \hline                       
 \end{tabular}
\caption{Properties of best-fitting (i.e. median) models and the 68\% confidence intervals for the low mass galaxies in our sample with the best UV spectra (see \S5.1 for details 
on modelling procedure).    In the bottom three 
rows, we present a subset of the line flux ratios and equivalent widths of the median model.  }
\end{table*}

\section{Photoionization modeling}

We now explore whether photoionisation models can reproduce the ultraviolet 
emission line ratios and equivalent widths of the dwarf galaxies.   First, we investigate 
whether the emission lines can be explained entirely 
by a low gas-phase metallicity, or if variations in the age, stellar metallicity, and  
ionisation parameter are also required.    Second, we examine whether there is 
enough energetic radiation output by low metallicity stellar populations to power the 
CIV emission seen in several dwarf systems.  

\subsection{Method}

We appeal to new models by Gutkin et al. (in preparation) to compute the emission from the photoionized interstellar gas in star-forming galaxies. 
These models are based on the prescription of \citet{charlot2001}, which combines a stellar population synthesis model with a photoionization code. 
In this approach, the ensemble of HII regions and the diffuse gas ionized by young stars throughout a galaxy are described by means of effective 
(i.e. galaxy-wide) parameters. The main adjustable parameters of the photoionized gas are the interstellar metallicity, $Z$, the typical ionization 
parameter of a newly ionized HII region, $U$ (which characterizes the ratio of ionizing-photon to gas densities at the edge of the Str\"{o}mgren sphere), 
and the dust-to-metal (mass) ratio, $\xi_{\mathrm d}$ (which characterizes the depletion of metals on to dust grains). 

The solar relative ratios of C, N, and O adopted by Gutkin et al. to reproduce the optical emission-line ratios of a large sample of nearby 
SDSS galaxies are $(\mathrm{C/O})_\odot\approx0.39$ and $(\mathrm{N/O})_\odot\approx0.09$. By default, C/O is assumed to not vary 
with gas metallicity, while Gutkin et al. appeal to the relation of \citet[][equation~5]{groves2004} to describe the dependence of N/O on 
metallicity. We also consider models with reduced C and N abundances at fixed metallicity, to describe the delayed release of C and N 
by intermediate-mass stars relative to shorter-lived massive stars, which are the main production sites of O \citep[and also produce C and N;][]{henry2000}
in young galaxies. For simplicity, we follow Erb et al. (2010) and consider models in which C and N are 
reduced in equal proportions relative to the default values at fixed metallicity, by factors of between 0.85 and 0.05.

We also include attenuation of line and continuum photons by dust in the neutral ISM, using the 2-component model of \citet{charlot2000}, 
as implemented by \citet[][their equations~1--4]{dacunha2008}. This is parameterized in terms of the total $V$-band attenuation optical depth of the dust, $\hat{\tau}_V$, and the fraction $\mu$ of this arising from dust in the diffuse ISM rather than in giant molecular clouds. Accounting for these two dust components is important to describe the attenuation of emission-line equivalent widths.

In their models, Gutkin et al. use the latest version of the \citet{bruzual2003} stellar population synthesis code to compute the emission 
from stars. Following \citet{charlot2001}, they neglect the contribution by stars older than 10\,Myr to nebular emission and use the 
latest version of the standard photoionization code CLOUDY \citep{ferland2013} to compute the emission-line spectrum generated 
by younger stars, assuming that galaxies are ionization bounded. In all calculations, the stellar metallicity is taken to be the 
same as that of the interstellar gas. The models considered here assume constant star formation rate and a standard 
\citet{chabrier2003} initial mass function. 

To interpret the nebular emission from the galaxies in our sample, we build a comprehensive grid of models covering a wide 
range of input parameters. Specifically, we take ionization parameters $\log U=-1.0$, $-1.5$, $-2.0$, $-2.5$, $-3.0$, $-3.5$, 
and $-4.0$; interstellar metallicities $Z=0.0001$, 0.0002, 0.0005, 0.001, 0.002, 0.004, 0.008, 0.017, and 0.03; dust-to-metal 
ratios $\xi_{\mathrm d}=0.1$, 0.3, and 0.5; C/O (and N/O) scaling factors 1.0, 0.85, 0.65, 0.45, 0.25, 0.15, and 0.05; about 
70 model ages between 10\,Myr and 1\,Gyr; 10 attenuation optical depths $\hat{\tau}_V$ between 0 and 1; and, for each 
$\hat{\tau}_V$, 10 values of $\mu$ between 0 and 1. We adopt a Bayesian approach similar to that 
of \citet[][their equation~1]{brinchmann2004} to compute the likelihood of each model given the data.     The  
reader is directed to this work for more detail.

\subsection{Results}

\begin{figure*}
\begin{center}
\includegraphics[width=0.48\textwidth, angle=270]{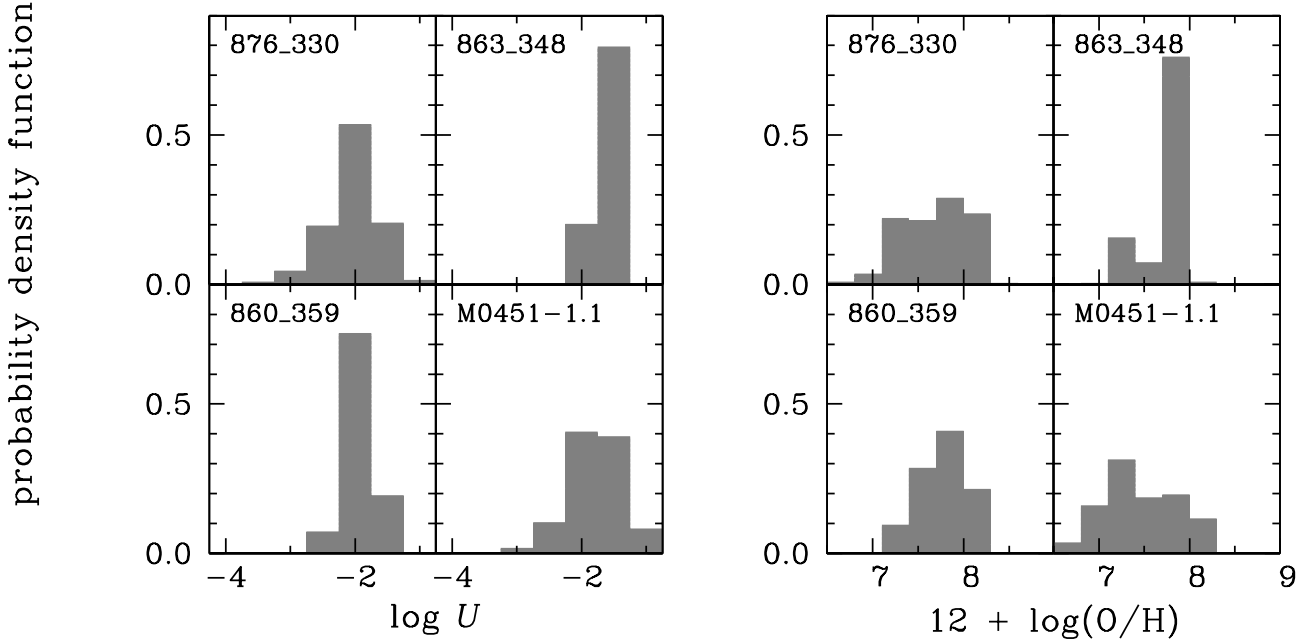}
\caption{Results from photoionisation modelling of CIII] emitters (\S5).   To reproduce the extreme ultraviolet equivalent widths, we require large 
ionisation parameters (left panel) and low metallicity gas and massive stars (right panel).   The models also require young ages (best fits 
of 6-51 Myr) and sub-solar carbon to oxygen abundances.   Details of the best fits are provided in Table 7.}
\label{fig:firespec}
\end{center}
\end{figure*}

We focus our modelling efforts on the four dwarf star forming galaxies with the highest quality 
spectra (876\_330, 863\_348, 860\_359, MACS 0451-1.1).    The CIII] equivalent widths spanned 
by this subset range from fairly typical values for dwarf star forming systems (6.7~\AA\ in MACS 0451-1.1) 
to the most extreme values in our sample (13.5~\AA\ in 863\_348).    As our goal is to determine what is 
required to produce the ultraviolet spectra, we do not fit the continuum SEDs or optical spectral 
features.   We instead find the range of models within our grid which provides the closest match to 
the equivalent width of CIII] and the flux ratios of CIV, OIII]$\lambda$1661, OIII]$\lambda$1666, 
Si III]$\lambda$1883, and Si III]$\lambda$1892 with respect to CIII].   We do not include the He II/CIII] ratio 
in the fitting process, as He II includes a stellar wind component which is non-trivial to decompose from 
the nebular component at low S/N (see Erb et al. 2010).  

The properties of the best fitting (i.e., median) models are listed in Table 7, revealing a very similar picture to 
that implied by the rest-optical spectra and broadband SEDs.    The extreme UV emission line strengths of our low mass 
galaxies strongly favour models with low oxygen abundances (0.04-0.13 Z$_\odot$), young ages (6-50 Myr) for an 
assumed constant star formation history, large ionisation parameters ($\log \rm{U}$ = $-2.16$ to $-1.84$), 
and low C/O ratios ($\log \rm{C/O}$ = $-0.74$ to $-0.58$).   The  probability density functions of $\log \rm{U}$ and 
$12+\log(\mathrm{O/H})$ are shown in Figure 8.    We note that as with the modelling of the continuum SED, the 
ages are only valid for the assumed star formation history.   More complicated star formation histories (i.e., those 
with a young and old component) would likely fit the data equally well as long as the recent star formation dominates the 
observed emission.   Given the ease in fitting the emission lines with a single component constant star formation 
history, we do not explore two component models.    As can be seen in Table 7, the models which best fit the ultraviolet 
spectra have large equivalent width [OIII]+H$\beta$ emission lines (600-1500 \AA\ rest-frame).   Not only are these 
values consistent with the observations in \S4.3, but they provide further evidence for a close connection 
between the optical EELG population and the strong UV line emitters in our 
sample.   

The first question we posed at the outset of this section was whether the prominent UV emission lines could be 
explained simply by the larger electron temperature that follows inefficient cooling in low metallicity gas.   
The best fitting parameters indicate that this is not the case.   Stellar population ages must be considerably lower and 
ionisation parameters somewhat larger than are commonly inferred for more massive star forming systems at high redshift (e.g., 
Hainline et al. 2009, Richard et al. 2011a, Reddy et al. 2012).    

The second question we motivated was whether the large equivalent width CIV emission 
seen in our most extreme galaxies could be powered by normal stellar populations without resorting to other 
energetic sources.   The CIV/CIII] ratios of the best fitting models (Table 7) indicate that moderately low metallicity 
models with large ionisation parameters produce ample energetic radiation to drive the observed CIV emission.  
We note that these same models predict  weak  He II$\lambda$1640 emission, in agreement with the 
observations presented in \S3.3.   While further data are certainly  required to determine if AGN, high mass X-ray binaries, or 
fast radiative shocks help power the high ionisation emission lines, the models suggest that they are not required to 
explain the CIV emission.   

\section{Discussion}

\subsection{New method of UV spectroscopy in reionization era}

In this paper, we have demonstrated that several ultraviolet emission lines (CIII], OIII], CIV)  
have particularly large equivalent widths in low mass star forming galaxies at high redshift.   
As a result of the attenuation of Ly$\alpha$ in $z\simeq 7$ galaxies (e.g., Schenker et al. 2012, Ono et al. 2012,
Pentericci et al. 2012, Treu et al. 2013, Schenker et al. 2014), these  UV emission lines likely provide our best hope 
of confirming very high redshift galaxies with ground-based facilities.   Even if new 
infrared spectrographs soon begin to yield more detections of Ly$\alpha$ at $z\gsim 7$, it will remain 
difficult to draw conclusions on the nature of early galaxies owing to resonant scattering of Ly$\alpha$ 
by neutral hydrogen in the IGM.   The ultraviolet emission emission lines described above provide an attractive alternative, opening 
the door for insight into the stellar populations and ionised gas conditions at $z\gsim 7$ through photoionisation 
modelling of line ratios and equivalent widths.  

Based on our understanding of the physical conditions and stellar populations that produce 
large equivalent width ultraviolet emission lines (as detailed in \S5.2), we 
predict that CIII] should be prominent in the spectra of $z\gsim 6$ galaxies.   The blue UV slopes of 
$z\gsim 6$ galaxies (e.g., Dunlop et al. 2013, Bouwens et al. 2013) are very similar to our sample of CIII] emitters 
at $z\simeq 2-3$ and likely reflect low metallicities and minimal reddening.    Once nebular emission contamination 
is accounted for at $z\gsim 6$, the specific star formation rates are found to be very large (Stark et al. 2013b, 
Smit et al. 2013, de Barros et al. 2013, Gonzalez et al. 2013), pointing to an ionising spectrum dominated 
by very massive stars.    And perhaps the 
most important smoking gun comes from recent inferences of large equivalent width rest-optical 
emission lines (Stark et al. 2013b, Smit et al. 2013, Labb\'{e} et al. 2013).   Based on the photoionisation 
models considered in \S5.2, extreme rest-frame optical emission lines are almost always accompanied by 
large equivalent width CIII] emission in the rest-frame ultraviolet.     
   
Detection of CIII] in the reionisation era need  not wait for future facilities.   For a bright (H$_{160}$=24.5) 
gravitationally-lensed galaxy at $z\simeq 6$, similar to many discovered recently (e.g., Richard et al. 2011b, Zitrin et al. 2012, 
Bradley et al. 2013), a CIII] emitter with rest-frame equivalent width between 7 and 13 \AA\ would have a  
line flux (3-6$\times$10$^{-18}$ erg cm$^{-2}$ s$^{-1}$) which is readily detectable with existing spectrographs.   
In the brightest known $z\gsim 6$ galaxies (J$_{\rm{125}}$=24.0 in Zitrin et al. 2012), 
an extreme CIII] emitter (similar to 863\_348) could 
have line fluxes as bright as 1.4$\times$10$^{-17}$ erg cm$^{-2}$ s$^{-1}$, requiring no more than an hour for detection.     

Because the CIII] doublet is likely to be resolved by most ground-based spectrographs at $z\gsim 6$, the 
total flux quoted above will be split among two components.   Integration times should be chosen to 
detect individual components.   Care must also be taken to avoid 
sources that place CIII] at wavelengths  affected by sky lines and atmospheric absorption.   Redshift between 5.2 and 6.1 
will have CIII] in the J-band, and between $z=6.9$ and $z=8.3$ the line will be in the H-band.    To  demonstrate the 
feasibility of detecting CIII] at $z\gsim 6$, it makes sense to first focus on the small subset of bright sources with spectroscopic 
confirmation from Ly$\alpha$.    Not only does the presence of large equivalent width Ly$\alpha$ increase confidence that CIII] will be 
present (based on Figure 4b), but for these sources, one can ensure that the observed wavelengths of CIII], OIII], and CIV 
are located in clean regions of the near-IR sky (after accounting for the small velocity offset between Ly$\alpha$ and systemic 
as discussed in \S3.4).   

Spectroscopic detection or robust limits on the equivalent widths of CIII], OIII], and CIV would provide new 
insight into the nature of $z\gsim 6$ galaxies.   If UV lines are detected, photoionisation modelling of the line strengths
(together with the broadband SED) 
will yield much improved constraints on the age and metallicity of the stellar populations, allowing us to put 
better constraints on the ionising output of galaxies in the reionisation era (e.g., Robertson et al. 2013).   In galaxies 
where Ly$\alpha$ is also present, detection of CIII] will provide a measure of the systemic redshift, allowing 
Ly$\alpha$ to be shifted into the rest-frame.   The derived Ly$\alpha$ velocity profile would provide new constraints  
on the transfer of Ly$\alpha$ through the outflowing neutral gas surrounding the galaxy.  By improving our understanding 
of how Ly$\alpha$ escapes from low mass galaxies at $z\gsim 6$, we will be better equipped to 
map Ly$\alpha$  evolution at $z\gsim 6$ to the ionisation state of the IGM.  

\subsection{Star formation in high redshift low mass galaxies}

If stellar feedback is  strong in low mass   
galaxies, star formation histories are likely to be strongly variable or bursty.  The imprint of such star formation fluctuations should be visible in the emission line and continuum properties of the lowest mass galaxies in 
our sample.  
In particular, the burstiness should cause the scatter in specific star formation rates to be greater in lower stellar mass 
galaxies.     A natural consequence of this is that there should be a significant population of  low mass galaxies 
undergoing very rapid stellar mass growth.   Such galaxies will be marked by large specific star formation 
rates and  very large equivalent width nebular emission lines.  The data we have collected suggest that 
extreme line emitters with large  ($\gsim 20$ Gyr$^{-1}$) specific star formation rates are fairly common among 
low luminosity star forming galaxies, similar to other recent results (e.g., van der Wel et al. 2011, Atek et al. 2011).   
The most extreme systems in our sample, such as 863\_348, require a hard 
radiation field from fairly young stellar populations ($\lsim 10$ Myr),  as might be expected for galaxies 
in the midst of a substantial upturn in their star formation.    

If star formation does fluctuate rapidly 
in low mass galaxies, then there  should also be a population with very low specific star 
formation rates.    If the period of low star formation activity lasts for $\gsim 10^7$ yr, 
such systems will lack the O and early B stars which power nebular emission lines.   In contrast, as long as 
there has been significant star formation activity in the past 100 Myr, the UV continuum (which is powered by O through later B 
type stars) will remain prominent.  
Both recombination and collisionally excited lines will thus be much weaker with respect to the UV continuum 
in this population.    Because our selection relies primarily on emission lines for redshift confirmation, our current sample is  
biased against finding low mass galaxies with low specific star formation rates.   But as we discussed in \S4.2 and \S4.4, we 
do identify one low stellar mass (5.1$\times$10$^6$ M$_\odot$) system, 869\_328, which might have recently undergone 
a downturn in its star formation.   While the sSFR derived from the UV continuum reflects very rapid stellar mass growth 
over the past 100 Myr, all indications suggest that nebular emission (line and continuum) are much weaker than for 
other systems with similar stellar continuum properties.   Either the region of the galaxy sampled by the slit 
has a large escape fraction of ionizing radiation, or 
the massive stars necessary to power nebular emission are absent owing to bursty star formation or a stochastic 
sampling of the IMF.   

Of course having observed only 17 low luminosity galaxies, a key question is whether the extreme UV and optical 
spectral features found in our sample are as common as implied by our data.    As discussed above, our spectroscopic 
selection certainly biases us toward objects with emission lines (although it should be noted that the spectra 
are of sufficient quality to detect 
absorption lines and low equivalent width line emission), and at the lowest stellar masses we are somewhat biased toward objects with 
large sSFR  because of our selection in the rest-frame UV.   In the coming years, WFC3/IR  observations 
of faint lensed galaxies in the HST Frontier Fields (GO: 13496, PI: Lotz, as well as through 
the  WFC3/IR grism survey GLASS, GO 13459; PI: Treu) should enable a more uniform census of the rest-frame optical nebular emission line properties of very low mass galaxies at high redshift, allowing a more reliable determination 
of whether there is substantially more scatter in the specific star formation rates at low masses.

 \section{Summary and future outlook}

We have presented observations of 17  low luminosity (M$_{\rm{UV}}$=$-13.7$ to $-19.9$) gravitationally-lensed galaxies at $z\simeq 1.5-3.0$.   
Stellar masses inferred from SED fitting range between 2.0$\times$10$^6$ M$_\odot$ and 1.4$\times$10$^9$ M$_\odot$ with a median 
of 6.3$\times$10$^{7}$ M$_\odot$.  Deep optical spectra  reveal prominent ultraviolet emission
 lines rarely seen in more luminous galaxies.   The blended 
CIII]$\lambda$1908 doublet is seen in 16 of 17 galaxies with an average equivalent width of 7.1~\AA, four times greater 
than that seen in the composite spectrum of luminous LBGs at $z\simeq 3$  (Shapley et al. 2003).   
Many of the galaxies also show fainter emission from OIII]$\lambda\lambda$1661,1666 and Si III]$\lambda\lambda$1883,1892. 
 The most extreme CIII] emitters in our sample have rest-frame equivalent widths as large as 13.5~\AA.\  These systems also show 
prominent emission from CIV$\lambda$1549, requiring a substantial flux of photons with energies greater than 47.9 eV, similar to 
the local population of blue compact dwarfs (e.g., Thuan \& Izotov 2005).  
Notably the nebular He II$\lambda$1640 emission line is weak or non-detected 
in many of the CIII] emitters.  We demonstrate that the equivalent width of CIII] is correlated with that of Ly$\alpha$ at $z\simeq 2-3$.    Galaxies with 
Ly$\alpha$ equivalent widths of greater than 50~\AA\ generally show CIII] emission with equivalent widths greater than 5~\AA.

With the goal of understanding the origin of the UV emission lines in our low mass galaxies, 
we have explored the physical properties in more detail in \S4.   While the star formation rates of 
our sample tend to be very low (median of 1.7 M$_\odot$ yr$^{-1}$),  the 
specific star formation rates are very large (median of 27 Gyr$^{-1}$), indicating that the galaxies 
are undergoing a period of rapid stellar mass growth.   The continuum UV slopes are very 
blue (median of $\beta = -2.2$), likely reflecting minimal dust content, young ages, and low metallicity.   The presence of powerful [OIII]+H$\beta$ emission is apparent from  the broadband SEDs, implying an average rest-frame equivalent width (770~\AA) similar 
to the population of EELGs identified in recent surveys.   Near-infrared spectra obtained with Magellan/FIRE confirm the presence of strong
rest-optical emission lines in a subset of our sample.   Oxygen abundances 
derived for extreme CIII] emitters indicate moderately metal poor gas ($\lsim 0.2$ Z$_\odot$), while 
the only object without CIII] emission appears to be somewhat more metal rich (0.4-0.5 Z$_\odot$).  

In \S5, we have considered whether the UV emission lines can be reproduced by a new suite of 
photoionisation models developed in Gutkin et al. (in preparation).    We find that the 
extreme CIII] emitting galaxies in our sample require models with large ionisation 
parameters (log U = $-2.16$ to $-1.84$), metal poor gas (0.04-0.13 Z$_\odot$), 
sub-solar C/O ratios (log C/O = -0.74 to -0.58), and a hard radiation field from moderately 
metal poor and young (6-50 Myr for constant star formation history) massive stars.   While the models suggest that AGN are not 
required to power the CIV emission lines, further observations are required to determine 
if additional heating sources are present.   

The data and models support a physical picture whereby the prominent UV emission lines 
arise from a confluence of factors.    The hard radiation 
field from young low metallicity stars increases the electron temperature in the ionised gas, which 
in turn increases the strength of collisionally-excited emission lines.   The large specific 
star formation rate indicates an enhanced ionising photon output per unit mass, increasing the equivalent width of nebular 
emission lines.   We suggest that the ubiquity of very large specific star formation rates 
in the low luminosity galaxies in our sample may reflect a recent upturn (or burst) in 
star formation.  Future work with more 
uniformly selected samples are required to further investigate the star formation histories 
implied by the ubiquity of large equivalent width emission lines in low mass galaxies at high redshift. 

Given the strong attenuation of Ly$\alpha$ in $z\gsim 6.5$ galaxies (e.g., Schenker et al. 2012, 
Pentericci et al. 2012, Ono et al. 2012, Treu et al. 2014, Schenker et al. 2014), we have argued that CIII] is likely to be the strongest UV spectral feature in $z\gsim 7$ galaxies.   Since ground-based studies are limited to the rest-UV at 
$z\gsim 6$, CIII] may provide our best  window on early galaxy formation in the era of 20-30 meter class telescopes.    
While the infrared spectroscopic capability of JWST will enable rest-optical lines to be detected to $z\simeq 8-10$, 
at yet higher redshifts, CIII], OIII], and CIV are likely our best hope 
for redshift confirmation and detailed study.    But detection of 
CIII] does not have to wait for future facilities.    We demonstrate that CIII] should be  
detectable in bright ($H\lsim 25$) gravitationally lensed galaxies now being located in 
HST imaging surveys (e.g., Richard et al. 2011b, Zitrin et al. 2012, Bradley et al. 2013).   In 
the brightest of these, multiple UV lines should be present, enabling new constraints on the 
stellar populations and ionised gas conditions in reionization-era galaxies. 

\section*{ACKNOWLEDGMENTS}

We thank the anonymous referee for a valuable report which strengthened 
the paper.    We also thank Richard Ellis, Dawn Erb, Max Pettini, and Alice Shapley 
for enlightening conversations.    WRF is grateful to Brad Holden for help with reduction of the 
Keck/LRIS data.   DPS acknowledges support from the National Science Foundation via grant AST-1410155.
JR acknowledges support from the European Research Council (ERC) starting grant CALENDS and 
the Marie Curie Career Integration Grant 294074.   
SC, JG and AW acknowledge support from the ERC via an Advanced Grant under grant agreement no. 321323ÑNEOGAL.  
RA acknowledge support from the Swedish Research Council and the Swedish National Space Board.
This paper includes data gathered with the 6.5 meter 
Magellan Telescopes located at Las Campanas 
Observatory, Chile.  Some of the data presented herein were obtained at the W.M. Keck Observatory, 
which is operated as a scientific partnership among the California Institute of Technology, the University 
of California and the National Aeronautics and Space Administration. The Observatory was made 
possible by the generous financial support of the W.M. Keck Foundation. Based on observations made with 
ESO telescopes at the La Silla Paranal Observatory under programme 088.A-0571.

\footnotesize{
  \bibliographystyle{mn2e}
  \bibliography{paper}

\begin{thebibliography}{}

\bibitem[Akerman et 
al.(2004)]{2004A&A...414..931A} Akerman, C.~J., Carigi, L., Nissen, P.~E., Pettini, M., \& Asplund, M.\ 2004, \aap, 414, 931 

\bibitem[Alavi et al.(2014)]{2014ApJ...780..143A} Alavi, A., Siana, B., 
Richard, J., et al.\ 2014, \apj, 780, 143 

\bibitem[Alexandroff et al.(2013)]{2013MNRAS.435.3306A} Alexandroff, R., 
Strauss, M.~A., Greene, J.~E., et al.\ 2013, \mnras, 435, 3306 

\bibitem[Amanullah et al.(2011)]{2011ApJ...742L...7A} Amanullah, R., 
Goobar, A., Cl{\'e}ment, B., et al.\ 2011, \apjl, 742, L7 

\bibitem[Amor{\'{\i}}n et al.(2014)]{Amorin14} Amor{\'{\i}}n, 
R., P{\'e}rez-Montero, E., Contini, T., et al.\ 2014, arXiv:1403.3441 

\bibitem[Atek et al.(2011)]{2011ApJ...743..121A} Atek, H., Siana, B., 
Scarlata, C., et al.\ 2011, \apj, 743, 121 

\bibitem[Atek et al.(2014)]{Atek14a} Atek, H., Kneib, J.-P., 
Pacifici, C., et al.\ 2014a, \apj, 789, 96 

\bibitem[Atek et al.(2014)]{Atek14b} Atek, H., Richard, J., 
Kneib, J.-P., et al.\ 2014b, \apj, 786, 60 

\bibitem[Atek et al.(2014)]{Atek14c} Atek, H., Kneib, J.-P., 
Pacifici, C., et al.\ 2014c, \apj, 789, 96 

\bibitem[Bayliss et al.(2013)]{2013arXiv1310.6695B} Bayliss, M.~B., Rigby, 
J.~R., Sharon, K., et al.\ 2013, submitted to ApJ arXiv:1310.6695 

\bibitem[Becker 
\& Bolton(2013)]{2013MNRAS.436.1023B} Becker, G.~D., \& Bolton, J.~S.\ 2013, \mnras, 436, 1023 

\bibitem[Behroozi et al.(2013)]{2013ApJ...770...57B} Behroozi, P.~S., 
Wechsler, R.~H., \& Conroy, C.\ 2013, \apj, 770, 57 

\bibitem[Bensby 
\& Feltzing(2006)]{2006MNRAS.367.1181B} Bensby, T., \& Feltzing, S.\ 2006, \mnras, 367, 1181 

\bibitem[Bouwens et al.(2012)]{2012ApJ...754...83B} Bouwens, R.~J., 
Illingworth, G.~D., Oesch, P.~A., et al.\ 2012, \apj, 754, 83 

\bibitem[Bouwens et al.(2013)]{2013arXiv1306.2950B} Bouwens, R.~J., 
Illingworth, G.~D., Oesch, P.~A., et al.\ 2013, submitted to ApJ, arXiv:1306.2950 

\bibitem[Bradley et al.(2013)]{2013arXiv1308.1692B} Bradley, L.~D., Zitrin, 
A., Coe, D., et al.\ 2013, arXiv:1308.1692 

\bibitem[Brammer et al.(2012)]{2012ApJ...758L..17B} Brammer, G.~B., 
S{\'a}nchez-Janssen, R., Labb{\'e}, I., et al.\ 2012, \apjl, 758, L17 

\bibitem[Brinchmann et al.(2004)]{brinchmann2004} Brinchmann, J., 
Charlot, S., White, S.~D.~M., et al.\ 2004, \mnras, 351, 1151 

\bibitem[Broadhurst et al.(2005)]{2005ApJ...621...53B} Broadhurst, T., 
Ben{\'{\i}}tez, N., Coe, D., et al.\ 2005, \apj, 621, 53 

\bibitem[Bruzual 
\& Charlot(2003)]{bruzual2003} Bruzual, G., \& Charlot, S.\ 2003, \mnras, 344, 1000 

\bibitem[Bunker et al.(2010)]{2010MNRAS.409..855B} Bunker, A.~J., Wilkins, 
S., Ellis, R.~S., et al.\ 2010, \mnras, 409, 855 

\bibitem[Calzetti et al.(2000)]{2000ApJ...533..682C} Calzetti, D., Armus, 
L., Bohlin, R.~C., et al.\ 2000, \apj, 533, 682 

\bibitem[Chabrier(2003)]{chabrier2003} Chabrier, G.\ 2003, \pasp, 
115, 763 

\bibitem[Charlot 
\& Fall(2000)]{charlot2000} Charlot, S., \& Fall, S.~M.\ 2000, \apj, 539, 718 

\bibitem[Charlot 
\& Longhetti(2001)]{charlot2001} Charlot, S., \& Longhetti, M.\ 2001, \mnras, 323, 887 


\bibitem[Christensen et al.(2012a)]{2012MNRAS.427.1953C} Christensen, L., 
Richard, J., Hjorth, J., et al.\ 2012a, \mnras, 427, 1953 

\bibitem[Christensen et al.(2012b)]{2012MNRAS.427.1973C} Christensen, L., 
Laursen, P., Richard, J., et al.\ 2012b, \mnras, 427, 1973 

\bibitem[Coe et al.(2014)]{Coe14} Coe, D., Bradley, L., 
\& Zitrin, A.\ 2014, arXiv:1405.0011 

\bibitem[Conroy 
\& Wechsler(2009)]{2009ApJ...696..620C} Conroy, C., \& Wechsler, R.~H.\ 2009, \apj, 696, 620 

\bibitem[Cooke et al.(2011)]{2011MNRAS.417.1534C} Cooke, R., Pettini, M., 
Steidel, C.~C., Rudie, G.~C., \& Nissen, P.~E.\ 2011, \mnras, 417, 1534 

\bibitem[Crowther et al.(2006)]{2006MNRAS.368..895C} Crowther, P.~A., 
Prinja, R.~K., Pettini, M., \& Steidel, C.~C.\ 2006, \mnras, 368, 895 

\bibitem[da Cunha et al.(2008)]{dacunha2008} da Cunha, E., Charlot, 
S., \& Elbaz, D.\ 2008, \mnras, 388, 1595 

\bibitem[de Barros et al.(2013)]{2013arXiv1207.3663D} de Barros, S., 
Schaerer, D., \& Stark, D.~P.\ 2013, arXiv:1207.3663 

\bibitem[Dekel 
\& Silk(1986)]{1986ApJ...303...39D} Dekel, A., \& Silk, J.\ 1986, \apj, 303, 39 

\bibitem[Diehl et al.(2009)]{2009ApJ...707..686D} Diehl, H.~T., Allam, 
S.~S., Annis, J., et al.\ 2009, \apj, 707, 686 

\bibitem[Dunlop et al.(2013)]{2013MNRAS.432.3520D} Dunlop, J.~S., Rogers, 
A.~B., McLure, R.~J., et al.\ 2013, \mnras, 432, 3520 

\bibitem[Efstathiou(1992)]{1992MNRAS.256P..43E} Efstathiou, G.\ 1992, 
\mnras, 256, 43P 

\bibitem[Eldridge et al.(2008)]{2008MNRAS.384.1109E} Eldridge, J.~J., 
Izzard, R.~G., \& Tout, C.~A.\ 2008, \mnras, 384, 1109 

\bibitem[Eldridge 
\& Stanway(2009)]{2009MNRAS.400.1019E} Eldridge, J.~J., \& Stanway, E.~R.\ 2009, \mnras, 400, 1019 

\bibitem[Eldridge 
\& Stanway(2012)]{2012MNRAS.419..479E} Eldridge, J.~J., \& Stanway, E.~R.\ 2012, \mnras, 419, 479 

\bibitem[Erb et al.(2010)]{erb2010} Erb, D.~K., Pettini, M., 
Shapley, A.~E., et al.\ 2010, \apj, 719, 1168 

\bibitem[Fabbian et 
al.(2009)]{2009A&A...500.1143F} Fabbian, D., Nissen, P.~E., Asplund, M., Pettini, M., \& Akerman, C.\ 2009, \aap, 500, 1143 

\bibitem[Ferland et al.(1998)]{1998PASP..110..761F} Ferland, G.~J., 
Korista, K.~T., Verner, D.~A., et al.\ 1998, \pasp, 110, 761 

\bibitem[Ferland et al.(2013)]{ferland2013} Ferland, G.~J., Porter, 
R.~L., van Hoof, P.~A.~M., et al.\ 2013, \rmxaa, 49, 137 

\bibitem[Finkelstein et al.(2012)]{2012ApJ...756..164F} Finkelstein, S.~L., 
Papovich, C., Salmon, B., et al.\ 2012, \apj, 756, 164 

\bibitem[F{\"o}rster Schreiber et al.(2009)]{2009ApJ...706.1364F} 
F{\"o}rster Schreiber, N.~M., Genzel, R., Bouch{\'e}, N., et al.\ 2009, 
\apj, 706, 1364 


\bibitem[Fosbury et al.(2003)]{2003ApJ...596..797F} Fosbury, R.~A.~E., 
Villar-Mart{\'{\i}}n, M., Humphrey, A., et al.\ 2003, \apj, 596, 797 

\bibitem[Garnett et al.(1995)]{1995ApJ...443...64G} Garnett, D.~R., 
Skillman, E.~D., Dufour, R.~J., et al.\ 1995, \apj, 443, 64 

\bibitem[Garnett et al.(1997)]{1997ApJ...481..174G} Garnett, D.~R., 
Skillman, E.~D., Dufour, R.~J., \& Shields, G.~A.\ 1997, \apj, 481, 174 

\bibitem[Garnett et al.(1999)]{1999ApJ...513..168G} Garnett, D.~R., 
Shields, G.~A., Peimbert, M., et al.\ 1999, \apj, 513, 168 

\bibitem[Gonzalez et al.(2012)]{2012arXiv1208.4362G} Gonzalez, V., Bouwens, 
R., llingworth, G., et al.\ 2012, arXiv:1208.4362 

\bibitem[Groves et al.(2004)]{groves2004} Groves, B.~A., Dopita, 
M.~A., \& Sutherland, R.~S.\ 2004, \apjs, 153, 9 

\bibitem[Guo et al.(2010)]{2010MNRAS.404.1111G} Guo, Q., White, S., Li, C., 
\& Boylan-Kolchin, M.\ 2010, \mnras, 404, 1111 

\bibitem[Guseva et al.(2000)]{2000ApJ...531..776G} Guseva, N.~G., Izotov, 
Y.~I., \& Thuan, T.~X.\ 2000, \apj, 531, 776 

\bibitem[Hainline et al.(2009)]{2009ApJ...701...52H} Hainline, K.~N., 
Shapley, A.~E., Kornei, K.~A., et al.\ 2009, \apj, 701, 52 

\bibitem[Hainline et al.(2011)]{2011ApJ...733...31H} Hainline, K.~N., 
Shapley, A.~E., Greene, J.~E., \& Steidel, C.~C.\ 2011, \apj, 733, 31 

\bibitem[Hashimoto et al.(2013)]{2013ApJ...765...70H} Hashimoto, T., Ouchi, 
M., Shimasaku, K., et al.\ 2013, \apj, 765, 70 

\bibitem[Henry et al.(2000)]{henry2000} Henry, R.~B.~C., Edmunds, 
M.~G., {\ Kouml}ppen, J.\ 2000, \apj, 541, 660

\bibitem[Holden et al.(2014)]{2014arXiv1401.5490H} Holden, B.~P., Oesch, 
P.~A., Gonzalez, V.~G., et al.\ 2014, arXiv:1401.5490 

\bibitem[Horne(1986)]{1986PASP...98..609H} Horne, K.\ 1986, \pasp, 98, 609  

\bibitem[Hu et al.(2009)]{2009ApJ...698.2014H} Hu, E.~M., Cowie, L.~L., 
Kakazu, Y., \& Barger, A.~J.\ 2009, \apj, 698, 2014 

\bibitem[James et al.(2013)]{2013arXiv1311.5092J} James, B.~L., Pettini, 
M., Christensen, L., et al.\ 2013, arXiv:1311.5092 

\bibitem[Jones et al.(2010)]{2010MNRAS.404.1247J} Jones, T.~A., Swinbank, 
A.~M., Ellis, R.~S., Richard, J., \& Stark, D.~P.\ 2010, \mnras, 404, 1247 

\bibitem[Jones et al.(2013)]{2013ApJ...765...48J} Jones, T., Ellis, R.~S., 
Richard, J., \& Jullo, E.\ 2013, \apj, 765, 48 

\bibitem[Jullo et al.(2007)]{2007NJPh....9..447J} Jullo, E., Kneib, J.-P., 
Limousin, M., et al.\ 2007, New Journal of Physics, 9, 447 

\bibitem[Kewley 
\& Ellison(2008)]{2008ApJ...681.1183K} Kewley, L.~J., \& Ellison, S.~L.\ 2008, \apj, 681, 1183 

\bibitem[Kewley 
\& Dopita(2002)]{2002ApJS..142...35K} Kewley, L.~J., \& Dopita, M.~A.\ 2002, \apjs, 142, 35 

\bibitem[Kewley et al.(2013)]{2013ApJ...774L..10K} Kewley, L.~J., Maier, 
C., Yabe, K., et al.\ 2013, \apjl, 774, L10 

\bibitem[Kewley et al.(2013)]{2013ApJ...774..100K} Kewley, L.~J., Dopita, 
M.~A., Leitherer, C., et al.\ 2013, \apj, 774, 100 

\bibitem[Kneib(1993)]{1993PhDT.......189K} Kneib, J.-P.\ 1993, 
Ph.D.~Thesis,  

\bibitem[Kobulnicky 
\& Kewley(2004)]{2004ApJ...617..240K} Kobulnicky, H.~A., \& Kewley, L.~J.\ 2004, \apj, 617, 240 

\bibitem[Kobulnicky 
\& Skillman(1998)]{1998ApJ...497..601K} Kobulnicky, H.~A., \& Skillman, E.~D.\ 1998, \apj, 497, 601 

\bibitem[Labb{\'e} et al.(2013)]{2013ApJ...777L..19L} Labb{\'e}, I., Oesch, 
P.~A., Bouwens, R.~J., et al.\ 2013, \apjl, 777, L19 

\bibitem[Lee et al.(2009)]{2009ApJ...706..599L} Lee, J.~C., Gil de Paz, A., 
Tremonti, C., et al.\ 2009, \apj, 706, 599 

\bibitem[Limousin et al.(2007)]{2007ApJ...668..643L} Limousin, M., Richard, 
J., Jullo, E., et al.\ 2007, \apj, 668, 643 

\bibitem[Maseda et al.(2013)]{2013ApJ...778L..22M} Maseda, M.~V., van der 
Wel, A., da Cunha, E., et al.\ 2013, \apjl, 778, L22 

\bibitem[Masters et al.(2014)]{2014arXiv1402.0510M} Masters, D., McCarthy, 
P., Siana, B., et al.\ 2014, arXiv:1402.0510 

\bibitem[McLure et al.(2013)]{2013MNRAS.432.2696M} McLure, R.~J., Dunlop, 
J.~S., Bowler, R.~A.~A., et al.\ 2013, \mnras, 432, 2696 

\bibitem[Moster et al.(2010)]{2010ApJ...710..903M} Moster, B.~P., 
Somerville, R.~S., Maulbetsch, C., et al.\ 2010, \apj, 710, 903 

\bibitem[Murray et al.(2005)]{2005ApJ...618..569M} Murray, N., Quataert, 
E., \& Thompson, T.~A.\ 2005, \apj, 618, 569 

\bibitem[Oesch et al.(2010)]{2010ApJ...725L.150O} Oesch, P.~A., Bouwens, 
R.~J., Carollo, C.~M., et al.\ 2010, \apjl, 725, L150 

\bibitem[Oesch et al.(2012)]{2012ApJ...759..135O} Oesch, P.~A., Bouwens, 
R.~J., Illingworth, G.~D., et al.\ 2012, \apj, 759, 135 

\bibitem[Osterbrock(1989)]{1989agna.book.....O} Osterbrock, D.~E.\ 1989, 
Research supported by the University of California, John Simon Guggenheim 
Memorial Foundation, University of Minnesota, et al.~Mill Valley, CA, 
University Science Books, 1989, 422 p.,  

\bibitem[Osterbrock 
\& Ferland(2006)]{2006agna.book.....O} Osterbrock, D.~E., \& Ferland, G.~J.\ 2006, Astrophysics of gaseous nebulae and active galactic nuclei, 2nd.~ed.~by D.E.~Osterbrock and G.J.~Ferland.~Sausalito, CA: University Science Books, 2006,  

\bibitem[Pacifici et al.(2012)]{2012MNRAS.421.2002P} Pacifici, C., Charlot, 
S., Blaizot, J., \& Brinchmann, J.\ 2012, \mnras, 421, 2002 

\bibitem[Papovich et al.(2001)]{2001ApJ...559..620P} Papovich, C., 
Dickinson, M., \& Ferguson, H.~C.\ 2001, \apj, 559, 620 

\bibitem[Pettini et al.(2000)]{2000ApJ...528...96P} Pettini, M., Steidel, 
C.~C., Adelberger, K.~L., Dickinson, M., 
\& Giavalisco, M.\ 2000, \apj, 528, 96 

\bibitem[Pettini 
\& Pagel(2004)]{2004MNRAS.348L..59P} Pettini, M., \& Pagel, B.~E.~J.\ 2004, \mnras, 348, L59 

\bibitem[Pontzen 
\& Governato(2012)]{2012MNRAS.421.3464P} Pontzen, A., \& Governato, F.\ 2012, \mnras, 421, 3464 

\bibitem[Price et al.(2013)]{2013arXiv1310.4177P} Price, S.~H., Kriek, M., 
Brammer, G.~B., et al.\ 2013, arXiv:1310.4177 

\bibitem[Quider et al.(2009)]{2009MNRAS.398.1263Q} Quider, A.~M., Pettini, 
M., Shapley, A.~E., \& Steidel, C.~C.\ 2009, \mnras, 398, 1263 

\bibitem[Quider et al.(2010)]{2010MNRAS.402.1467Q} Quider, A.~M., Shapley, 
A.~E., Pettini, M., Steidel, C.~C., 
\& Stark, D.~P.\ 2010, \mnras, 402, 1467 

\bibitem[Raiter et 
al.(2010)]{2010A&A...523A..64R} Raiter, A., Schaerer, D., \& Fosbury, R.~A.~E.\ 2010, \aap, 523, A64 

\bibitem[Reddy 
\& Steidel(2009)]{2009ApJ...692..778R} Reddy, N.~A., \& Steidel, C.~C.\ 2009, \apj, 692, 778 

\bibitem[Reddy et al.(2012)]{2012ApJ...744..154R} Reddy, N., Dickinson, M., 
Elbaz, D., et al.\ 2012a, \apj, 744, 154 

\bibitem[Reddy et al.(2012)]{2012ApJ...754...25R} Reddy, N.~A., Pettini, 
M., Steidel, C.~C., et al.\ 2012b, \apj, 754, 25 

\bibitem[Richard et al.(2007)]{2007ApJ...662..781R} Richard, J., Kneib, 
J.-P., Jullo, E., et al.\ 2007, \apj, 662, 781 

\bibitem[Richard et al.(2011)]{2011MNRAS.413..643R} Richard, J., Jones, T., 
Ellis, R., et al.\ 2011a, \mnras, 413, 643 

\bibitem[Richard et al.(2011)]{2011MNRAS.414L..31R} Richard, J., Kneib, 
J.-P., Ebeling, H., et al.\ 2011b, \mnras, 414, L31 

\bibitem[Rix et al.(2004)]{2004ApJ...615...98R} Rix, S.~A., Pettini, M., 
Leitherer, C., et al.\ 2004, \apj, 615, 98 

\bibitem[Robertson et al.(2010)]{2010Natur.468...49R} Robertson, B.~E., 
Ellis, R.~S., Dunlop, J.~S., McLure, R.~J., 
\& Stark, D.~P.\ 2010, \nat, 468, 49 

\bibitem[Robertson et al.(2013)]{2013ApJ...768...71R} Robertson, B.~E., 
Furlanetto, S.~R., Schneider, E., et al.\ 2013, \apj, 768, 71 

\bibitem[Rogers et al.(2013)]{2013MNRAS.429.2456R} Rogers, A.~B., McLure, 
R.~J., \& Dunlop, J.~S.\ 2013, \mnras, 429, 2456 

\bibitem[Salpeter(1955)]{1955ApJ...121..161S} Salpeter, E.~E.\ 1955, \apj, 
121, 161 

\bibitem[Schaerer(2003)]{2003A&A...397..527S} Schaerer, D.\ 2003, \aap, 397, 527 

\bibitem[Schenker et al.(2012)]{2012ApJ...744..179S} Schenker, M.~A., 
Stark, D.~P., Ellis, R.~S., et al.\ 2012, \apj, 744, 179 

\bibitem[Schenker et al.(2013a)]{2013ApJ...768..196S} Schenker, M.~A., 
Robertson, B.~E., Ellis, R.~S., et al.\ 2013a, \apj, 768, 196 

\bibitem[Schenker et al.(2013b)]{2013ApJ...777...67S} Schenker, M.~A., 
Ellis, R.~S., Konidaris, N.~P., \& Stark, D.~P.\ 2013b, \apj, 777, 67 

\bibitem[Schenker et al.(2014)]{2014arXiv1404.4632S} Schenker, M.~A., 
Ellis, R.~S., Konidaris, N.~P., \& Stark, D.~P.\ 2014, submitted to ApJ, arXiv:1404.4632 

\bibitem[Schmidt et al.(2014)]{2014ApJ...782L..36S} Schmidt, K.~B., Treu, 
T., Brammer, G.~B., et al.\ 2014, \apjl, 782, L36 

\bibitem[Shapley et al.(2003)]{2003ApJ...588...65S} Shapley, A.~E., 
Steidel, C.~C., Pettini, M., \& Adelberger, K.~L.\ 2003, \apj, 588, 65 

\bibitem[Shapley et al.(2005)]{2005ApJ...626..698S} Shapley, A.~E., 
Steidel, C.~C., Erb, D.~K., et al.\ 2005, \apj, 626, 698 

\bibitem[Shapley(2011)]{2011ARA&A..49..525S} Shapley, A.~E.\ 2011, \araa, 49, 525 

\bibitem[Shen et al.(2013)]{2013arXiv1308.4131S} Shen, S., Madau, P., 
Conroy, C., Governato, F., \& Mayer, L.\ 2013, submitted to ApJ, arXiv:1308.4131 

\bibitem[Shim et al.(2011)]{2011ApJ...738...69S} Shim, H., Chary, R.-R., 
Dickinson, M., et al.\ 2011, \apj, 738, 69 

\bibitem[Shirazi 
\& Brinchmann(2012)]{2012MNRAS.421.1043S} Shirazi, M., \& Brinchmann, J.\ 2012, \mnras, 421, 1043 


\bibitem[Simcoe et al.(2013)]{2013PASP..125..270S} Simcoe, R.~A., 
Burgasser, A.~J., Schechter, P.~L., et al.\ 2013, \pasp, 125, 270 


\bibitem[Smit et al.(2013)]{2013arXiv1307.5847S} Smit, R., Bouwens, R.~J., 
Labbe, I., et al.\ 2013, arXiv:1307.5847 

\bibitem[Smith et al.(2005)]{2005MNRAS.359..417S} Smith, G.~P., Kneib, 
J.-P., Smail, I., et al.\ 2005, \mnras, 359, 417 

\bibitem[Stark et al.(2009)]{2009ApJ...697.1493S} Stark, D.~P., Ellis, 
R.~S., Bunker, A., et al.\ 2009, \apj, 697, 1493 

\bibitem[Stark et al.(2010)]{2010MNRAS.408.1628S} Stark, D.~P., Ellis, 
R.~S., Chiu, K., Ouchi, M., \& Bunker, A.\ 2010, \mnras, 408, 1628 

\bibitem[Stark et al.(2011)]{2011ApJ...728L...2S} Stark, D.~P., Ellis, 
R.~S., \& Ouchi, M.\ 2011, \apjl, 728, L2 

\bibitem[Stark et al.(2013a)]{2013ApJ...763..129S} Stark, D.~P., Schenker, 
M.~A., Ellis, R., et al.\ 2013a, \apj, 763, 129 

\bibitem[Stark et al.(2013b)]{2013MNRAS.436.1040S} Stark, D.~P., Auger, M., 
Belokurov, V., et al.\ 2013b, \mnras, 436, 1040 

\bibitem[Steidel et al.(2010)]{2010ApJ...717..289S} Steidel, C.~C., Erb, 
D.~K., Shapley, A.~E., et al.\ 2010, \apj, 717, 289 

\bibitem[Stinson et al.(2007)]{2007ApJ...667..170S} Stinson, G.~S., 
Dalcanton, J.~J., Quinn, T., Kaufmann, T., 
\& Wadsley, J.\ 2007, \apj, 667, 170 

\bibitem[Tapken et 
al.(2007)]{2007A&A...467...63T} Tapken, C., Appenzeller, I., Noll, S., et al.\ 2007, \aap, 467, 63 

\bibitem[Thuan 
\& Izotov(2005)]{2005ApJS..161..240T} Thuan, T.~X., \& Izotov, Y.~I.\ 2005, \apjs, 161, 240 

\bibitem[Treu et al.(2013)]{2013ApJ...775L..29T} Treu, T., Schmidt, K.~B., 
Trenti, M., Bradley, L.~D., \& Stiavelli, M.\ 2013, \apjl, 775, L29 

\bibitem[van der Wel et al.(2011)]{2011ApJ...742..111V} van der Wel, A., 
Straughn, A.~N., Rix, H.-W., et al.\ 2011, \apj, 742, 111 

\bibitem[Vanzella et 
al.(2010)]{2010A&A...513A..20V} Vanzella, E., Grazian, A., Hayes, M., et al.\ 2010, \aap, 513, A20 

\bibitem[Vernet et 
al.(2011)]{2011A&A...536A.105V} Vernet, J., Dekker, H., D'Odorico, S., et al.\ 2011, \aap, 536, A105 


\bibitem[Wild et al.(2011)]{2011MNRAS.417.1760W} Wild, V., Charlot, S., 
Brinchmann, J., et al.\ 2011, \mnras, 417, 1760 


\bibitem[Wilkins et al.(2011)]{2011MNRAS.417..717W} Wilkins, S.~M., Bunker, 
A.~J., Stanway, E., Lorenzoni, S., \& Caruana, J.\ 2011, \mnras, 417, 717 


\bibitem[Wuyts et al.(2011)]{2011ApJ...738..106W} Wuyts, S., F{\"o}rster 
Schreiber, N.~M., Lutz, D., et al.\ 2011, \apj, 738, 106 


\bibitem[Yuan et al.(2013)]{2013ApJ...767..106Y} Yuan, T.-T., Kewley, 
L.~J., \& Rich, J.\ 2013, \apj, 767, 106 

\bibitem[Zitrin et al.(2012)]{2012ApJ...747L...9Z} Zitrin, A., Moustakas, 
J., Bradley, L., et al.\ 2012, \apjl, 747, L9 

\end{thebibliography}
}

\label{lastpage}

\end{document}